\documentclass[twocolumn]{aastex63}

\usepackage{amsmath}

\newcommand{\msun}{\hbox{M$_{\odot}$}}

\newcommand{\halpha}{\hbox{H$\alpha$}}
\newcommand{\hbeta}{\hbox{H$\beta$}}

\newcommand{\swift}{\textit{Swift }}

\graphicspath{{./}{figures/}}

\begin{document}

\title{The Curious Case of ASASSN-20hx: A Slowly-Evolving, UV and X-ray Luminous, Ambiguous Nuclear Transient}
\shorttitle{The Ambiguous Nuclear Transient ASASSN-20hx}

\submitjournal{ApJ}

\shortauthors{Hinkle et al.}

\correspondingauthor{Jason T. Hinkle}
\email{jhinkle6@hawaii.edu}

\author[0000-0001-9668-2920]{Jason T. Hinkle}
\affiliation{Institute for Astronomy, University of Hawai`i, 2680 Woodlawn Drive, Honolulu, HI 96822, USA}

\author[0000-0001-9206-3460]{Thomas W.-S. Holoien}
\altaffiliation{NHFP Einstein Fellow}
\affiliation{The Observatories of the Carnegie Institution for Science, 813 Santa Barbara Street, Pasadena, CA 91101, USA}

\author[0000-0003-4631-1149]{Benjamin. J. Shappee}
\affiliation{Institute for Astronomy, University of Hawai`i, 2680 Woodlawn Drive, Honolulu, HI 96822, USA}

\author[0000-0001-7351-2531]{Jack M.~M.~Neustadt}
\affiliation{Department of Astronomy, The Ohio State University, 140 West 18th Avenue, Columbus, OH 43210, USA}

\author[0000-0002-4449-9152]{Katie~Auchettl}
\affiliation{School of Physics, The University of Melbourne, Parkville, VIC 3010, Australia}
\affiliation{ARC Centre of Excellence for All Sky Astrophysics in 3 Dimensions (ASTRO 3D)}
\affiliation{Department of Astronomy and Astrophysics, University of California, Santa Cruz, CA 95064, USA}

\author{Patrick J. Vallely}
\affiliation{Department of Astronomy, The Ohio State University, 140 West 18th Avenue, Columbus, OH 43210, USA}

\author[0000-0002-9301-5302]{Melissa Shahbandeh}
\affiliation{Department of Physics, Florida State University, 77 Chieftan Way, Tallahassee, FL 32306, USA}

\author{Matthias Kluge}
\affiliation{University-Observatory, Ludwig-Maximilians-University, Scheinerstrasse 1, D-81679 Munich, Germany}
\affiliation{Max Planck Institute for Extraterrestrial Physics, Giessenbachstrasse, D-85748 Garching, Germany}

\author[0000-0001-6017-2961]{Christopher S. Kochanek}
\affiliation{Department of Astronomy, The Ohio State University, 140 West 18th Avenue, Columbus, OH 43210, USA}
\affiliation{Center for Cosmology and Astroparticle Physics, The Ohio State University, 191 W.~Woodruff Avenue, Columbus, OH 43210, USA}

\author{K. Z. Stanek}
\affiliation{Department of Astronomy, The Ohio State University, 140 West 18th Avenue, Columbus, OH 43210, USA}
\affiliation{Center for Cosmology and Astroparticle Physics, The Ohio State University, 191 W.~Woodruff Avenue, Columbus, OH 43210, USA}

\author[0000-0003-1059-9603]{Mark E. Huber}
\affiliation{Institute for Astronomy, University of Hawai`i, 2680 Woodlawn Drive, Honolulu, HI 96822, USA}

\author{Richard S. Post}
\affiliation{Post Observatory, Lexington, MA 02421, USA}

\author{David Bersier}
\affiliation{Astrophysics Research Institute, Liverpool John Moores University, 146 Brownlow Hill, Liverpool L3 5RF, UK}

\author{Christopher Ashall}
\affiliation{Institute for Astronomy, University of Hawai`i, 2680 Woodlawn Drive, Honolulu, HI 96822, USA}

\author[0000-0002-2471-8442]{Michael A. Tucker}
\altaffiliation{DOE CSGF Fellow}
\affiliation{Institute for Astronomy, University of Hawai`i, 2680 Woodlawn Drive, Honolulu, HI 96822, USA}

\author[0000-0001-5058-695X]{Jonathan P. Williams}
\affiliation{Institute for Astronomy, University of Hawai`i, 2680 Woodlawn Drive, Honolulu, HI 96822, USA}


\author{Thomas de Jaeger}
\affiliation{Institute for Astronomy, University of Hawai`i, 2680 Woodlawn Drive, Honolulu, HI 96822, USA}

\author{Aaron Do}
\affiliation{Institute for Astronomy, University of Hawai`i, 2680 Woodlawn Drive, Honolulu, HI 96822, USA}

\author{Michael Fausnaugh}
\affiliation{Department of Physics and Kavli Institute for Astrophysics and Space Research, Massachusetts Institute of Technology, Cambridge, MA 02139, USA}

\author{Daniel Gruen}
\affiliation{University-Observatory, Ludwig-Maximilians-University, Scheinerstrasse 1, D-81679 Munich, Germany}
\affiliation{Kavli Institute for Particle Astrophysics \& Cosmology, P. O. Box 2450, Stanford University, Stanford, CA 94305, USA}

\author[0000-0003-1008-225X]{Ulrich Hopp}
\affiliation{University-Observatory, Ludwig-Maximilians-University, Scheinerstrasse 1, D-81679 Munich, Germany}
\affiliation{Max Planck Institute for Extraterrestrial Physics, Giessenbachstrasse, D-85748 Garching, Germany}

\author{Justin Myles}
\affiliation{Department of Physics, Stanford University, 382 Via Pueblo Mall, Stanford, CA 94305, USA}

\author{Christian Obermeier}
\affiliation{University-Observatory, Ludwig-Maximilians-University, Scheinerstrasse 1, D-81679 Munich, Germany}
\affiliation{Max Planck Institute for Extraterrestrial Physics, Giessenbachstrasse, D-85748 Garching, Germany}

\author[0000-0003-3490-3243]{Anna V. Payne}
\altaffiliation{NASA Fellowship Activity Fellow}
\affiliation{Institute for Astronomy, University of Hawai`i, 2680 Woodlawn Drive, Honolulu, HI 96822, USA}

\author{Todd A. Thompson}
\affiliation{Department of Astronomy, The Ohio State University, 140 West 18th Avenue, Columbus, OH 43210, USA}
\affiliation{Center for Cosmology and Astroparticle Physics, The Ohio State University, 191 W.~Woodruff Avenue, Columbus, OH 43210, USA}

\begin{abstract}
We present observations of ASASSN-20hx, a nearby ambiguous nuclear transient (ANT) discovered in NGC 6297 by the All-Sky Automated Survey for Supernovae (ASAS-SN).  We observed ASASSN-20hx from $-$30 to 275 days relative to peak UV/optical emission using high-cadence, multi-wavelength spectroscopy and photometry. From Transiting Exoplanet Survey Satellite (TESS) data, we determine that the ANT began to brighten on 2020 June 22.8 with a linear rise in flux for at least the first week. ASASSN-20hx peaked in the UV/optical 30 days later on 2020 July 22.8 (MJD = 59052.8) at a bolometric luminosity of $L = (3.15 \pm 0.04) \times 10^{43}$ erg s$^{-1}$. The subsequent decline is slower than any TDE observed to date and consistent with many other ANTs. Compared to an archival X-ray detection, the X-ray luminosity of ASASSN-20hx increased by an order of magnitude to $L_{x} \sim 1.5 \times 10^{42}$ erg s$^{-1}$ and then slowly declined over time. The X-ray emission is well-fit by a power law with a photon index of $\Gamma \sim 2.3 - 2.6$. Both the optical and near infrared spectra of ASASSN-20hx lack emission lines, unusual for any known class of nuclear transient. While ASASSN-20hx has some characteristics seen in both tidal disruption events (TDEs) and active galactic nuclei (AGNs), it cannot be definitively classified with current data.
\end{abstract}


\section{Introduction}

All massive galaxies are known to host supermassive black holes (SMBHs) in their centers \citep[e.g.,][]{magorrian98, kormendy13}. These SMBHs play a critical role in moderating galaxy evolution through radiative and mechanical feedback \citep[e.g.,][]{kormendy13}. Direct detections of inactive SMBHs are  limited to our own SMBH \citep[e.g.,][]{ghez05} or massive SMBHs in nearby galaxies due to the need to resolve the SMBH’s small sphere of influence \citep[e.g.,][]{ford94, atkinson05, gebhardt11}. Nonetheless, $\sim1-5$\% of galaxies in the local universe host actively accreting SMBHs \citep{haggard10, lacerda20}, easily identified as active galactic nuclei (AGNs) \citep{kauffmann03}.

While the steady-state properties of AGNs provide the opportunity for detailed study across the electromagnetic spectrum \citep[e.g.,][]{sanders88, boroson92, reynolds97}, AGNs are also known to vary in a stochastic manner. This includes both photometric \citep[e.g.,][]{macleod12} and spectroscopic \citep[e.g.,][]{bianchi05} variability. Perhaps the most striking examples of AGN variability are the so-called changing-look AGNs \citep[e.g.,][]{shappee14, denney14, macleod16, yang18} characterised by the appearance or disappearance of broad emission lines over time. In addition to long-timescale changes in activity over several years, AGNs also undergo dramatic flaring behavior \citep[e.g.,][]{drake09, graham17, trakhtenbrot19b, frederick20} on much shorter timescales. In addition to flares occurring in unambiguous AGNs, there is a growing class of rapid turn-on events \citep[e.g.,][]{gezari17, gromadzki19, trakhtenbrot19a, neustadt20, frederick19}, characterised by the rapid appearance of a blue continuum and strong emission lines in galaxies that do not host obvious AGNs.

Beyond AGN activity, the tidal disruption of a star and the subsequent return of debris to the SMBH results in a luminous accretion flare \citep[][]{rees88, phinney89, evans89, ulmer99}. These tidal disruption events (TDEs) can occur in any galaxy and are thus good probes of otherwise inactive SMBHs. While early theoretical work predicted that the TDE spectral energy distribution (SED) should peak in the soft X-ray band \citep[e.g.][]{lacy82, rees88, evans89, phinney89}, observational studies have discovered that TDE properties are quite varied \citep[e.g.,][]{vanvelzen11, holoien14b, holoien16a, hinkle20a, alexander20, vanvelzen21}. This includes a broad diversity in spectroscopic properties \citep{holoien19b, leloudas19, hung20, vanvelzen21}, the existence or lack of radio emission \citep{alexander20}, and the presence or lack of X-ray emission \citep{holoien16a, auchettl17, wevers20}.

Optically-selected TDEs show luminous UV/optical emission which is well-fit by a blackbody with a temperature on the order of $20,000-50,000$ K \citep[e.g.,][]{gezari12b, holoien14b, holoien16a}. TDE blackbody temperatures are often roughly constant throughout their evolution, although some objects experience temperature evolution \citep{holoien19a, vanvelzen21, hinkle20a}. This is in contrast to the UV/optical continuum emission from AGNs, which is well-fit by a power law \citep[e.g.,][]{vandenberk01}. Some TDEs also have soft X-ray emission \citep[e.g.,][]{holoien16a, holoien16b, wevers19, hinkle21a}, which can generally be described by a blackbody with kT $\sim30-60$ eV \citep{auchettl17}. Conversely, X-ray emission from AGNs is ubiquitous \citep[e.g.,][]{mushotzky93} and harder than typical TDE X-ray emission \citep[e.g.,][]{ricci17, auchettl18}. AGN X-ray spectra are consistent with Comptonization \citep[e.g.,][]{poutanen96} and often exhibit hard X-rays above $\sim10$ keV, far more energetic than X-rays usually seen for TDEs.

The photometric evolution of TDEs and AGN-related flares also show drastic differences. The light curves of TDEs often exhibit a smooth and monotonic decline \cite[e.g.,][]{holoien14b, holoien19a, holoien20, nicholl20, hinkle21a, vanvelzen21}. In particular, the high-cadence Transiting Exoplanet Survey Satellite \citep[TESS;][]{ricker15} light curve of the TDE ASASSN-19bt \citep{holoien19b} illustrates the smoothness of TDE light curves. This is in stark contrast to the stochastic variability superimposed on some AGN flares \cite[e.g.,][]{neustadt20, frederick19, frederick20} and the existence of AGN flares with rebrightening episodes \citep[e.g.,][]{frederick20, malyali21}. In addition, the masses of SMBHs hosting TDEs occur are relatively low \citep[M $\la 10^{7.5}$ M$\odot$][]{wevers17, vanvelzen18} because the tidal radius for a main sequence star around a higher mass SMBH lies within the event horizon \citep[e.g.,][]{rees88}.

The spectra of TDEs and AGN flares are also quite distinct. Optical TDE spectra are blue, often with very broad ($\ga 10000$ km s$^{-1}$) H or He lines \citep[e.g.,][]{arcavi14, holoien16a, holoien16b, vanvelzen21}. Some TDEs also exhibit emission from metal lines \citep[e.g.,][]{blagorodnova17, leloudas19} or Bowen fluorescence features \citep[e.g.,][]{blagorodnova17, leloudas19, vanvelzen21}. These are in contrast to the typical spectra of AGN, which have emission from Balmer lines and forbidden lines such as [\ion{O}{3}] and [\ion{N}{2}]. The line widths seen in AGN spectra during outburst are relatively narrow, on the order of $\sim 2000$ km s$^{-1}$ \citep{frederick20}. Additionally, the UV spectra of TDEs and AGNs are distinct due to differences in the environments in which they occur. AGNs exhibit strong [\ion{Mg}{2}] emission in the UV \citep{vandenberk01, batra14}, which is conspicuously absent in all TDE UV spectra \citep{brown18, hung21}. Interestingly, the line evolution of TDEs and AGNs is also quite  different. In most TDEs, there is a positive correlation between line luminosity and line width \citep[e.g.,][]{holoien16a, holoien19a, hinkle21a}. AGNs show the opposite behavior, with emission line widths increasing as the luminosity decreases \citep[e.g.,][]{peterson04, denney09}.

Despite their many differences, AGNs and TDEs both provide a detailed look at accretion physics \citep[e.g.,][]{merloni03, lodato11, guillochon15}, and insight into the environment and growth of SMBHs \citep[e.g.][]{antonucci93, auchettl18}. Additionally, the observed emission from both TDEs and AGNs is sensitive to SMBH parameters like spin and mass \citep[e.g.][]{reynolds14, mockler19, gafton19, reynolds19}. Because of the low fraction of AGNs in local galaxies, comprehensive studies of both AGNs and TDEs are needed to provide complete picture of such properties of SMBHs and galactic nuclei.

Fortunately, with the advent of transient surveys like the All-Sky Automated Survey for Supernovae \citep[ASAS-SN;][]{shappee14, kochanek17}, the Asteroid Terrestrial Impact Last Alert System \citep[ATLAS;][]{tonry18}, the Zwicky Transient Facility \citep[ZTF;][]{bellm19}, and the Panoramic Survey Telescope and Rapid Response System \citep[Pan-STARRS;][]{chambers16} many more TDEs and AGN flares are being discovered. In addition, ground-based surveys have begun to discover an increasing number of ambiguous nuclear transients (ANTs), which cannot be neatly classified into either source class. Examples of such objects include ASASSN-18jd \citep{neustadt20}, ASASSN-18el \citep{trakhtenbrot19b, ricci20}, and the ANT studied here, ASASSN-20hx. These sources represent a unique opportunity to probe the variety of SMBH triggering mechanisms and search for broad trends among SMBH accretion behaviors.

In this paper we present the discovery and observations of ASASSN-20hx. This paper is organized as follows. In Section \ref{obs} we detail the discovery and observations of the ANT. In Section \ref{analysis} we present our analysis. Section \ref{disc} provides a discussion of our results and finally, we summarize our findings in Section \ref{summary}. Throughout the paper we assume a cosmology of $H_0$ = 69.6 km s$^{-1}$ Mpc$^{-1}$, $\Omega_{M} = 0.29$, and $\Omega_{\Lambda} = 0.71$.

\section{Discovery and Observations}\label{obs}

ASASSN-20hx $(\alpha,\delta =$ 17:03:36.49, $+$62:01:32.34) was discovered in the $g$-band data from the ASAS-SN ``Brutus'' unit on Haleakal\={a}, Hawai`i on 2020 July 10.3 UTC \citep{brimacombe20}. Its discovery was announced on the Transient Name Server (TNS), and assigned the name AT 2020ohl\footnote{\url{https://wis-tns.weizmann.ac.il/object/2020ohl}}. Rather than anonymize the discovering survey, we will continue to refer to the ANT by its public survey name ASASSN-20hx. ASASSN-20hx is located in the nucleus of the elliptical galaxy NGC 6297, a nearby S0 galaxy \citep{devacouleurs91} at a redshift of z = 0.01671 \citep{adelmanmccarthy06}, corresponding to a luminosity distance of 72.9 Mpc. \citet{saulder06} find a distance to NGC 6297 of 55.6 Mpc using fundamental plane relationships, but due to the large scatter in the fundamental plane, we will continue to use a distance of 72.9 Mpc throughout the remainder of this work.

Shortly after discovery, we obtained several spectroscopic observations. Spectra from both the SPectrograph for the Rapid Acquisition of Transients \citep[SPRAT;][]{piascik14} on the 2-m Liverpool Telescope \citep{steele04} and the Low-Resolution Imaging Spectrometer \citep[LRIS;][]{oke95} on the 10-m Keck I telescope, showed a blue continuum with few strong spectral features compared to the Sloan Digital Sky Survey \citep[SDSS;][]{york00} host spectrum (see \S 2.1). The strong blue continuum and a position consistent with the nucleus of the host galaxy made ASASSN-20hx a potential TDE candidate. Based on this, we triggered spectroscopic and ground-based photometric follow-up of ASASSN-20hx.

From these spectra and several \textit{Neil Gehrels Swift Gamma-ray Burst Mission} \citep[\textit{Swift};][]{gehrels04} observations, \citet{hinkle20_atel} classified the source as a TDE based on the observations of multiple blue spectra, a hot blackbody temperature, position in the center of the host galaxy, and the lack of spectral features usually associated with AGNs or supernovae. However, the X-ray emission from ASASSN-20hx is harder than a typical TDE \citep{lin20} and the subsequent evolution of ASASSN-20hx was unlike any known TDE, suggesting it is a more exotic transient.

\subsection{Archival Data of NGC 6297} \label{archival}
NGC 6297 has been observed by several surveys across the electromagnetic spectrum. We obtained $ugriz$ and $JHK_S$ images from SDSS Data Release 15 \citep{aguado19} and the Two Micron All-Sky Survey \citep[2MASS;][]{skrutskie06, 2MASS_img}, respectively. We measured aperture magnitudes using a 15\farcs{0} aperture radius in order to capture all of the galaxy light, and used several stars in the field to calibrate the magnitudes. We also measured a 15\farcs{0} aperture $NUV$ magnitude from the Galaxy Evolution Explorer \citep[GALEX;][]{martin05} images using gPhoton \citep{million16}. Finally, we obtained archival $W1$, $W2$, $W3$, and $W4$ magnitudes from the Wide-field Infrared Survey Explorer \citep[WISE;][]{wright10} AllWISE catalog \citep{allwise_doi, cutri14}, giving us coverage from ultraviolet through mid-infrared wavelengths.

\begin{figure}
\centering
 \includegraphics[width=.48\textwidth]{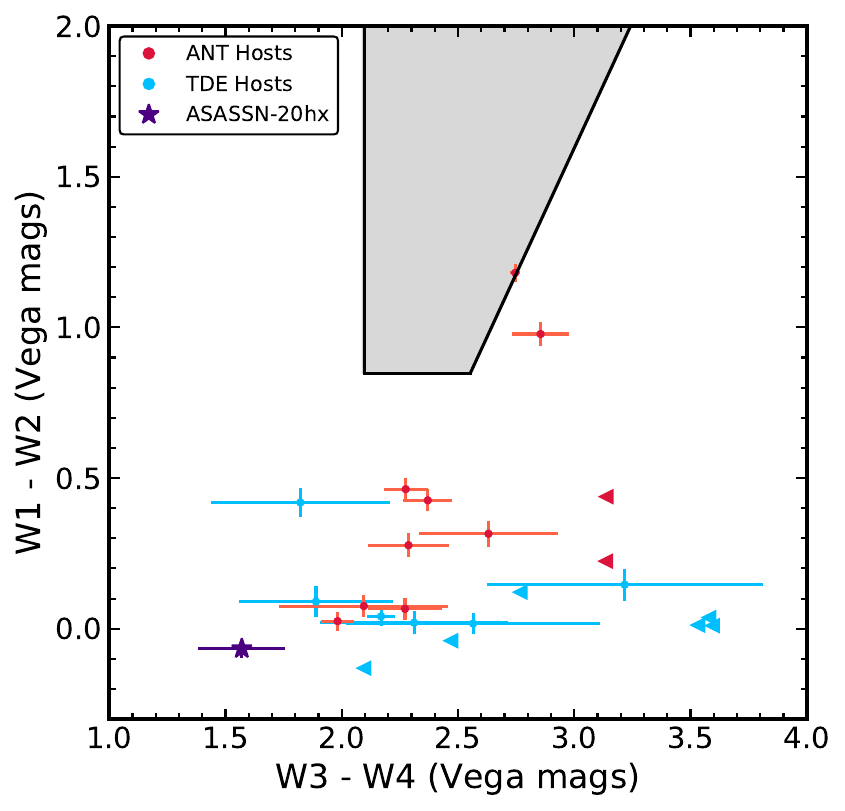}
 \caption{WISE color-color diagram used to discriminate strong AGNs from star-forming galaxies. The two colors used are $W1-W2$ and $W3-W4$ incorporating each of the WISE filters. The gray box is the AGN region determined by \citet{assef13}. In addition to the host galaxy of ASASSN-20hx (purple point) we compare to other ANT hosts (red points) and TDE hosts (blue points) from the sample of \citet{hinkle21b}. Like ASASSN-20hx, few of these hosts are selected as a strong AGN.}
 \label{fig:wise}
\end{figure}

\begin{deluxetable}{ccc}
\tablewidth{240pt}
\tabletypesize{\footnotesize}
\tablecaption{Archival Host Galaxy Photometry}
\tablehead{
\colhead{Filter} &
\colhead{AB Magnitude} &
\colhead{Magnitude Uncertainty}}
\startdata
$NUV$ & 19.01 & 0.05 \\
$u$ & 16.03 & 0.06 \\
$g$ & 14.28 & 0.02 \\
$r$ & 13.55 & 0.02 \\
$i$ & 13.20 & 0.02 \\
$z$ & 12.90 & 0.02 \\
$J$ & 12.61 & 0.02 \\
$H$ & 12.40 & 0.03 \\
$K_S$ & 12.59 & 0.03 \\
$W1$ & 13.65 & 0.02 \\
$W2$ & 14.35 & 0.02 \\
\enddata 
\tablecomments{Archival magnitudes of the host galaxy NGC 6297 used for our SED. modeling. NUV, $ugriz$, and $JHK_S$ magnitudes are 15\farcs{0} aperture magnitudes measured from GALEX, SDSS and 2MASS images, respectively. The $W1$ and $W2$ magnitudes are taken from the WISE AllWISE catalog. All magnitudes are presented in the AB system.} 
\label{tab:arch_phot} 
\end{deluxetable}

In order to constrain the possibility of the host galaxy being an AGN, we analyzed a range of archival data for NGC 6297. From an archival Swift X-ray Telescope (XRT; \citealt{burrows05}) epoch taken $\sim385$ days prior to discovery, we find weak X-ray emission from the host galaxy at a level of $0.012 \pm 0.007$ counts s$^{-1}$. Assuming an AGN with a photon index of $\Gamma = 1.75$ \citep[e.g.,][]{ricci17} and a Galactic column density of $N_{H}=2.0\times10^{20}$ cm$^{-2}$ along the line of sight \citep{HI4PI16}, this corresponds to an unabsorbed flux of $(5.0\pm 3.0) \times 10^{-13} \text{ erg } \text{cm}^{-2} \text{ s}^{-1}$ in the 0.3 - 10 keV band. At the distance of NGC 6297, this yields an X-ray luminosity of $(3.4 \pm 2.1)  \times 10^{41} \text{ erg } \text{ s}^{-1}$. A detection at this level rules out strong AGN activity, and is consistent with a weak or low luminosity AGN \citep[LLAGN; ][]{tozzi06, marchesi16, liu17, ricci17}. We also measured $UVW1$ and $U$ magnitudes from archival \swift images taken as part of the Swift Gravitational Wave Galaxy Survey \citep[SGWGS;][]{evans12}. We find that over the span of roughly a year NGC 6297 brightened from $U = 16.86 \pm 0.04$ mag and $UVW1 = 18.46 \pm 0.08$ mag on MJD = 58326.4 to $U = 16.71 \pm 0.04$ mag and $UVW1 = 18.06 \pm 0.08$ mag on MJD = 58655.1. As the increased near-UV magnitudes are coincident with the weak X-ray detection, this adds additional support to the statement that NGC 6297 may be a LLAGN.

The MIR colors of the host, ($W1-W2$) = $-0.07 \pm 0.03$ mag and ($W3-W4$) = $1.57 \pm 0.19$ mag, also suggest that NGC 6297 does not harbour an AGN as luminous as the host galaxy \citep[e.g.,][]{assef13}. As shown in Figure \ref{fig:wise}, the host galaxy of ASASSN-20hx lies far from the AGN region and is bluer than most TDE and ANT hosts, consistent with light predominantly from stars. The NeoWISE mission \citep{mainzer11, NEOWISE_phot} $W1$ and $W2$ band light curves show weak evidence for variability as the reduced $\chi^2$ values for fitting them as a constant are 2.6 and 2.3 respectively.

We fit stellar population synthesis models to the archival photometry of NGC 6297 (shown in Table~\ref{tab:arch_phot}) using the Fitting and Assessment of Synthetic Templates \citep[\textsc{Fast};][]{kriek09} to obtain an SED of the host. Our fit assumes a \citet{cardelli88} extinction law with $\text{R}_{\text{V}} = 3.1$, Galactic extinction of $\text{A}_{\text{V}} = 0.064$ mag \citep{schlafly11}, a Salpeter IMF \citep{salpeter55}, an exponentially declining star-formation rate, and the \citet{bruzual03} stellar population models. Based on the \textsc{Fast} fit, NGC 6297 has a stellar mass of M$_* = (4.2_{-0.4}^{+0.5}) \times 10^{10}$~\msun, an age of $(4.7_{-0.5}^{+0.7})$ Gyr, and star formation rate of SFR $(2.3_{-0.2}^{+0.9}) \times 10^{-2}$ M$_{\odot}$ yr$^{-1}$. Using the sample of \citet{mendel14} to compute a scaling relation between stellar mass and bulge mass, we estimate a bulge mass of $\sim 10^{10.5}$~\msun. We then use the M$_B$ - M$_{BH}$ relation of \citet{mcconnell13} to estimate a black hole mass of $\sim 10^{7.9}$~\msun. This SMBH mass is higher than for most TDEs \citep{kochanek16b, vanvelzen18, ryu20}.

\begin{deluxetable}{ccccc}
\tablewidth{240pt}
\tabletypesize{\footnotesize}
\tablecaption{Synthetic Host-Galaxy Magnitudes}
\tablehead{
\colhead{Filter} &
\colhead{AB Mag} &
\colhead{Mag Unc.} &
\colhead{Observed Mag} &
\colhead{Mag Unc.}}
\startdata
$UVW2$ & 19.94 & 0.19 & --- & --- \\
$UVM2$ & 19.69 & 0.13 & --- & ---  \\
$UVW1$ & 18.69 & 0.10 & 18.31 & 0.06 \\
$u$ & 16.57 & 0.06 & 16.71 & 0.08 \\
$U$ & 16.68 & 0.06 & 16.79 & 0.03 \\
$B$ & 15.32 & 0.04 & --- & ---  \\
$g$ & 15.01 & 0.03 & 14.95 & 0.03 \\
$V$ & 14.49 & 0.02 & --- & ---  \\
$r$ & 14.19 & 0.02 & 14.25 & 0.02 \\
$i$ & 13.87 & 0.02 & 13.94 & 0.02 \\
$z$ & 13.55 & 0.02 & 13.54 & 0.02 \\
$J$ & 13.22 & 0.02 & 13.25 & 0.02 \\
$H$ & 12.98 & 0.02 & 13.02 & 0.02 \\
\enddata  
\tablecomments{Synthetic host magnitudes of NGC 6297 from our \textsc{FAST} SED fits to the 5\farcs{0} magnitudes. These are used to subtract the fluxes of our follow-up photometry for which archival magnitudes are unavailable. The right columns indicated the observed archival magnitudes and their uncertainties. The archival \swift $UVW1$ and $U$ magnitudes come from stacking the two SGWGS epochs. All magnitudes are presented in the AB system.} 
\label{tab:synth_phot} 
\end{deluxetable}

Our photometric follow-up campaign includes several filters for which archival imaging data are not available, including the majority of the \swift UVOT and $BVRI$ filters. In comparing our \swift and ground-based $BV$ photometry, we found a better agreement when using the default \swift 5\farcs{0} aperture radius rather than the larger 15\farcs{0} radius. We therefore re-measured 5\farcs{0} aperture magnitudes of the host galaxy in the $NUV$, $ugriz$ and $JHK$ filters. We again fit stellar population synthesis models to this data using \textsc{Fast}. While this does not incorporate light from the entire host galaxy, we nevertheless find good agreement between the SFRs and ages estimated from the 15\farcs{0} and 5\farcs{0} fits. In order to estimate the host flux in these filters for host subtraction, we convolved the host SED from \textsc{Fast} with the filter response curve for each filter to obtain 5\farcs{0} fluxes. To estimate uncertainties on the estimated host galaxy fluxes, we perturbed the archival host fluxes assuming Gaussian errors and ran 1000 different \textsc{Fast} iterations. We present these 5\farcs{0} synthetic magnitudes in Table \ref{tab:synth_phot}. These synthetic fluxes were then used to subtract the host flux in our non-survey follow-up data.

\begin{figure*}
\centering
 \includegraphics[width=.48\textwidth]{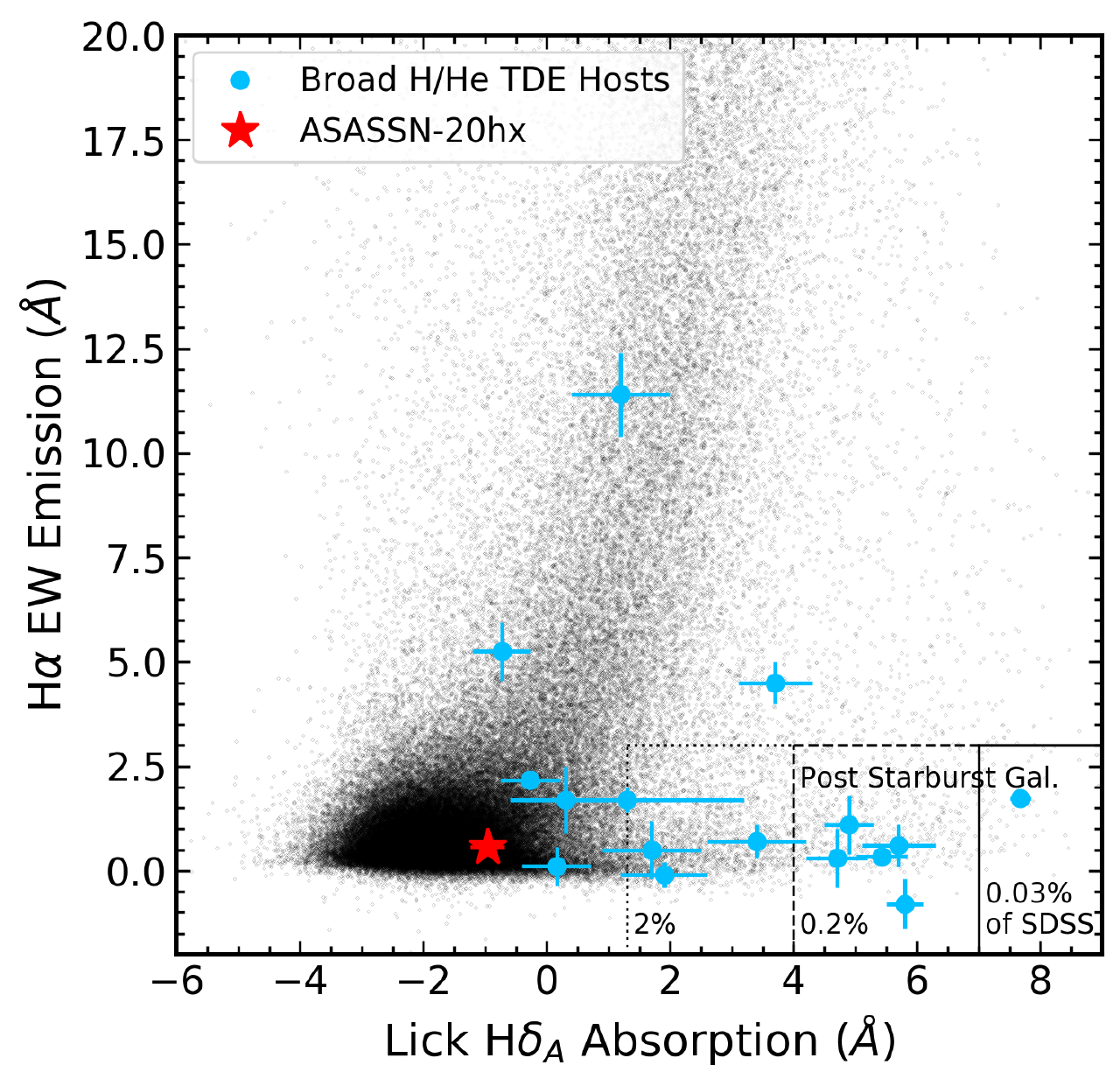}\hfill
 \includegraphics[width=.48\textwidth]{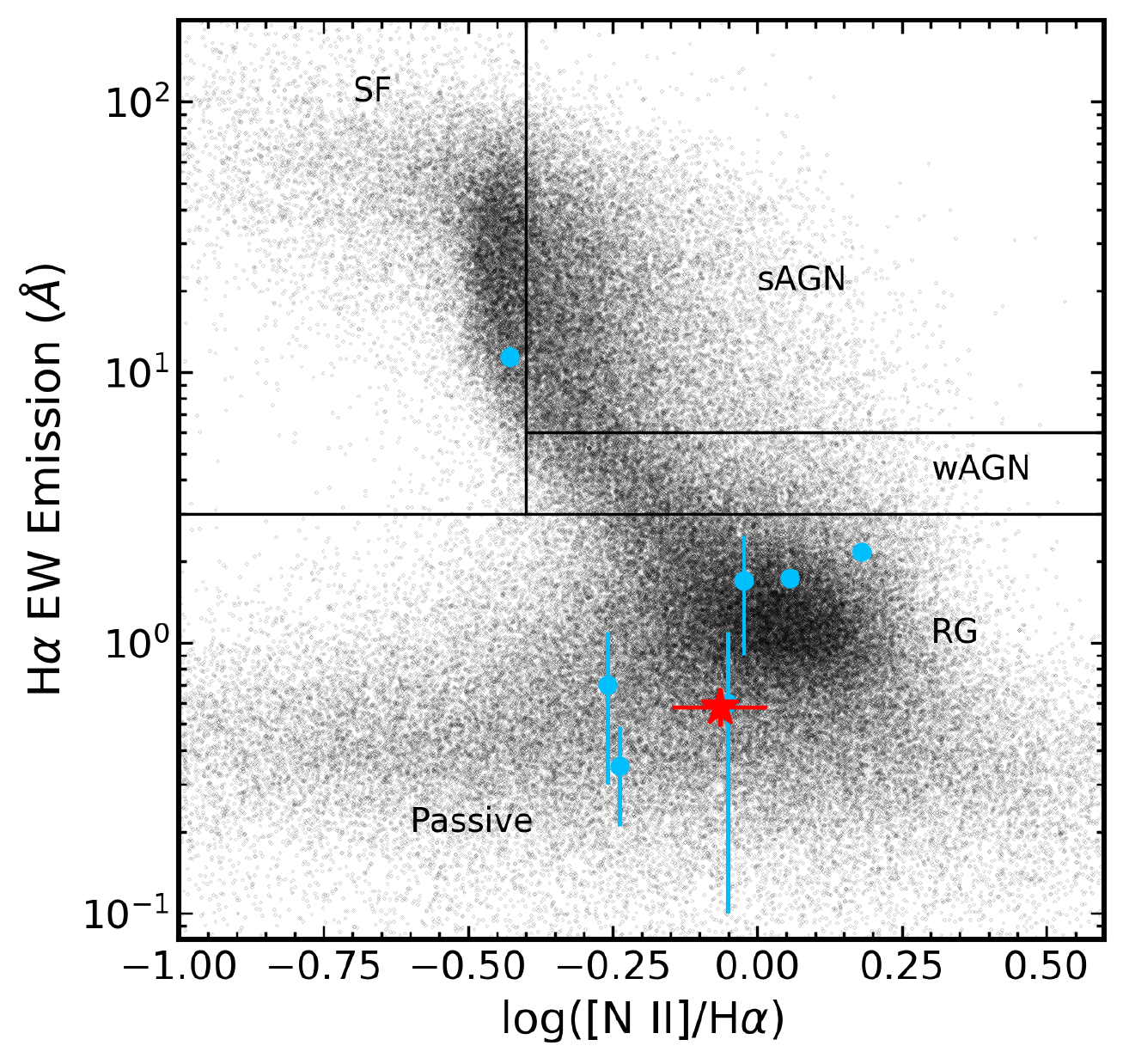} \\
 \caption{\textit{Left Panel}: H$\alpha$ emission line equivalent width (EW), which traces current star formation, as compared to the Lick H$\delta_A$ absorption index, which traces star formation in the past Gyr. The host galaxy NGC 6297 is shown as a red star, with TDE hosts shown as blue circles. NGC 6297 lies in the large cloud of points with little current or recent star formation. The error bars on NGC 6297 are roughly the size of the symbol. \textit{Right Panel}: H$\alpha$ emission line equivalent width (W$_{H\alpha}$), as compared to log$_{10}$([\ion{N}{2}] / \halpha), otherwise known as the WHAN diagram \citep{cidfernandes11}. Lines separating star-forming galaxies (SF), strong AGN (sAGN), weak AGN (wAGN), and passive and ``retired galaxies'' (RG) are shown \citep{cidfernandes11}. In all panels, galaxies from SDSS Data Release 8 \citep{eisenstein11} are shown in black.}
 \label{fig:ew_bpt}
\end{figure*}

We also examined the archival SDSS \citep{york00} spectrum of NGC 6297. This spectrum shows [\ion{N}{2}] $\lambda$6584 and weak [\ion{O}{3}] $\lambda\lambda$4959, 5007 and [\ion{S}{2}] $\lambda\lambda$6717, 6731 emission with Balmer lines in absorption. Figure \ref{fig:ew_bpt} shows the properties of NGC 6297 in several strong emission line diagnostic diagrams using the line fluxes from the MPA-JHU catalog \citep{brinchmann04}. The left panel of Figure \ref{fig:ew_bpt} compares the H$\alpha$ emission line equivalent width to the Lick H$\delta_A$ absorption index, which compares current and past star formation to identify post-starburst galaxies \citep[e.g.,][]{french16}. The right panel of Figure \ref{fig:ew_bpt} shows H$\alpha$ emission equivalent width as compared to log$_{10}$([\ion{N}{2}]$/$\halpha) to separate ionization mechanisms, particularly those associated with LINER-like (Low-Ionization Nuclear Emission-line Region) emission line ratios. The background points show the general distribution of galaxies in the MPA-JHU catalog \citep{brinchmann04} for SDSS DR8 \citep{eisenstein11}. 

From Figure \ref{fig:ew_bpt}, we see that NGC 6297 lies in the large cloud of points corresponding to galaxies without active star formation, consistent with the SED modeling. In fact, the Lick H$\delta_A$ absorption index is lower than most of the TDE host galaxies. The WHAN (H$\alpha$ equivalent width vs.\ [\ion{N}{2}]$/$\halpha) diagram of Figure \ref{fig:ew_bpt} places NGC 6297 near the ``retired galaxies'' (RG) region, where galaxies have ceased actively forming stars and are predominantly ionized by hot low-mass evolved stars such as post-AGB stars \citep{cidfernandes11}. However, the host lies slightly outside the cloud of RG points and towards the passive galaxies region, supported by the lack of current star formation in the galaxy.

While the SDSS spectrum shows only weak emission lines, affecting typical BPT analysis, we include the commonly-studied emission line ratio measurements here for completeness. The line ratios log$_{10}$([\ion{O}{3}]$/$\hbeta) = $0.33 \pm 0.13$, log$_{10}$([\ion{N}{2}]$/$\halpha) = $-0.07 \pm 0.08$, and log$_{10}$([\ion{S}{2}]$/$\halpha) = $-0.44 \pm 0.21$ place this galaxy in the AGN region of the ([\ion{N}{2}]$/$\halpha \ diagram and the \ion{H}{2} region of the [\ion{S}{2}]$/$\halpha \ diagram \citep{baldwin81, veilleux87}. Given this disagreement and the large uncertainties, the dominant ionization mechanism is uncertain in this galaxy. Combinations of star formation, AGN activity, and shocks, are known to produce intermediate line ratios such as those seen for NGC 6297 \citep[e.g.,][]{davies14, dagostino19}. We note that the fraction of detected Seyferts and LINERs in S0 galaxies is $\sim12$\% and $\sim26$\% respectively \citep{ho08}. This strengthens the case that NGC 6297 may host a LLAGN.

\begin{figure*}
\centering
 \includegraphics[width=\textwidth]{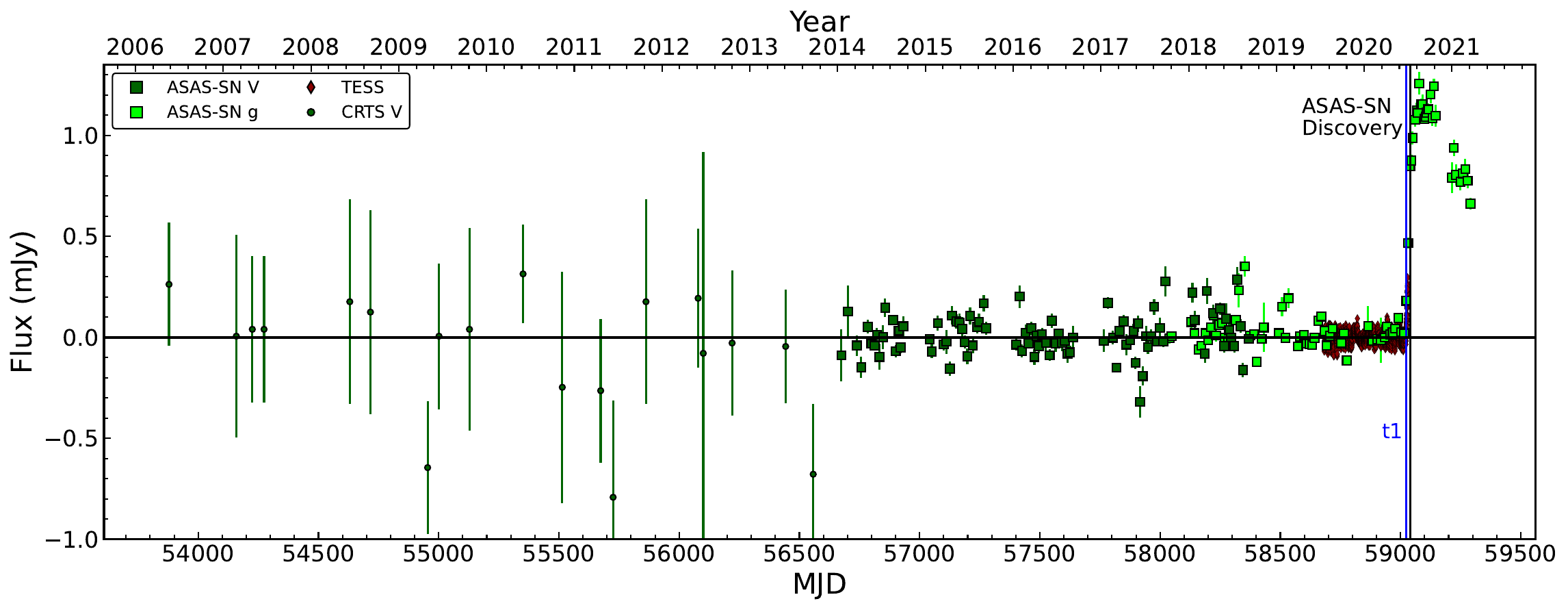}
 \caption{Long-term host-subtracted light curve of NGC 6297 from CRTS, ASAS-SN, and TESS. Dark green circles show CRTS $V$-band data stacked in 50-day bins. Squares represent ASAS-SN photometry, with dark green showing $V$-band and light green showing $g$-band data. The ASAS-SN data is stacked in 10-day bins prior to the flare and 5 day bins after the flare. Red diamonds show TESS data in 2-hour bins. The horizontal black line represents zero flux. The vertical blue line is the time of first light obtained from our fit to the TESS data and the vertical black line  marks the time of the ASAS-SN discovery.}
 \label{fig:long_lc}
\end{figure*}

In addition to our spectroscopic analysis of the host galaxy of ASASSN-20hx, we also looked for nuclear variability using archival Catalina Real-Time Transient Survey \citep[CRTS;][]{drake09}, ASAS-SN, and TESS. CRTS uses SExtractor \citep[][]{bertin96} photometry, which both includes flux from the host galaxy and makes the light curve noisy. We simply subtracted the mean flux as an estimate of the host contamination. These light curves are shown in Figure \ref{fig:long_lc}. No significant variability is seen over this timescale.

\subsection{ASAS-SN Light Curve}
ASAS-SN is a fully automated transient survey consisting of 20 individual telescopes on 5 robotic mounts. Each telescope is a 14-cm aperture Nikon telephoto lens with 8\farcs{0} pixels, and each unit consists of 4 telescopes sharing a common mount. The five ASAS-SN units are located at Haleakal\=a Observatory, McDonald Observatory, the South African Astrophysical Observatory (SAAO), and two at Cerro Tololo Inter-American Observatory (CTIO). With this current telescope network, ASAS-SN monitors the visible sky with a cadence of $\sim 20$ hours to a depth of $g \sim 18.5$ mag.

The images were reduced using the automated ASAS-SN pipeline, which incorporates the ISIS image subtraction package \citep{Alard1998, Alard2000}. As we have several years of ASAS-SN $g$-band photometry prior to the ASASSN-20hx outburst, we built a reference image using good images from multiple cameras observed at least a month before the fit time of first light. This common reference image was then scaled subtracted from all the $g$-band data. The ASAS-SN $V$-band data was reduced using the default ASAS-SN pipeline and references as there is no transient flux seen in any $V$-band epoch.

We then used the IRAF {\tt apphot} package with a 2-pixel radius (approximately 16\farcs{0}) aperture to perform aperture photometry on each subtracted image, generating a differential light curve. Our photometry was calibrated using the AAVSO Photometric All-Sky Survey \citep{Henden2015}. We discarded images with a FWHM of 1.67 pixels or greater. We stacked the pre-flare ASAS-SN data in 10-day bins to get deep early-time limits.

\subsection{\textit{TESS} Observations}

The host galaxy of ASASSN-20hx, NGC 6297, lies in the TESS northern continuous viewing zone (CVZ) near the North Ecliptic Pole, so it was observed continuously between 2019-Jul-18 and 2020-Jul-04. While the ASASSN-20hx flare was only observed in Sector 26 at the very end of Cycle 2, we nevertheless have almost a full year of high-precision photometry to search for  prior variability and study the onset of the flare.

Similar to our ASAS-SN data reduction, we used the ISIS package \citep{alard98, alard00} to perform image subtraction on the TESS full frame images (FFIs) to produce light curves \citep[see ][]{vallely19, vallely21, fausnaugh21}. Because of TESS's large pixel scale, we constructed independent reference images for each sector. To do this, we selected the first 100 FFIs of good quality obtained in each sector, excluding images with sky background levels or PSF widths above average for the sector. We excluded all FFIs with data quality flags. We also used several conservative quality cuts such as excluding FFIs obtained when the spacecraft’s pointing was compromised, when TESS was affected by an instrument anomaly, or when significant background effects due to scattered light were present in the images.

\begin{figure}
\centering
 \includegraphics[width=0.49\textwidth]{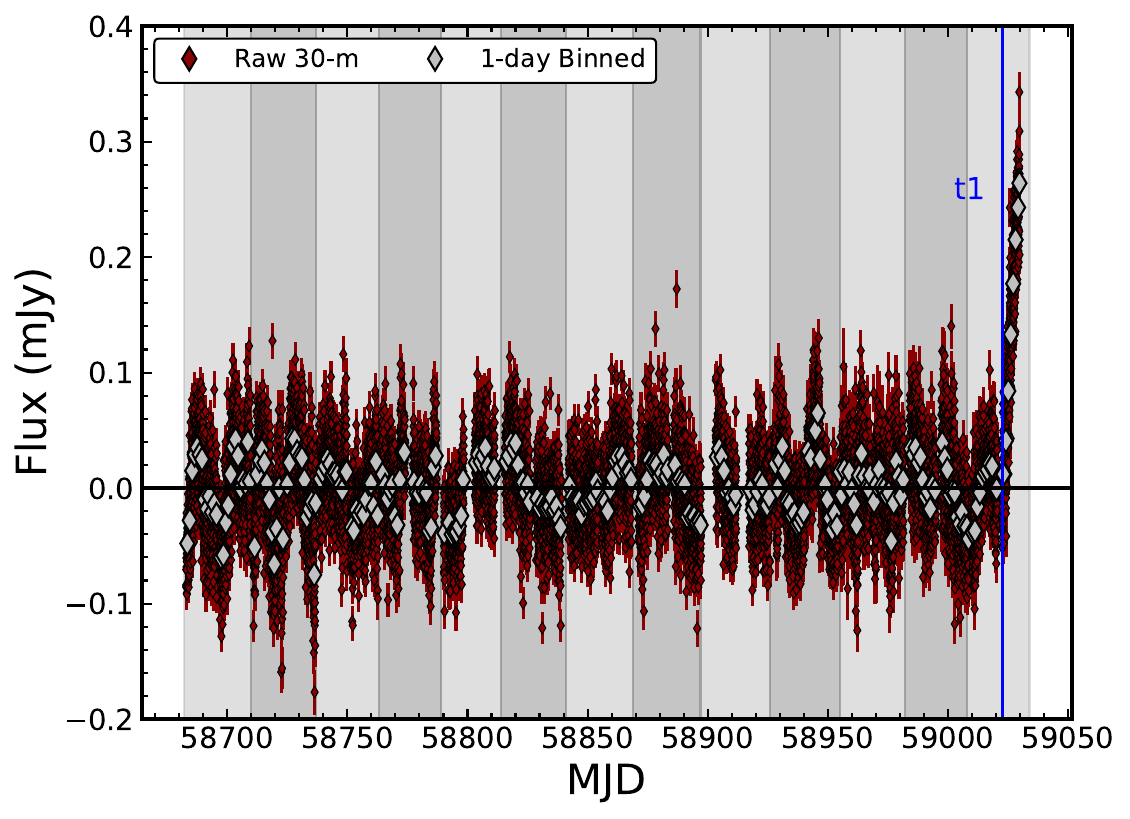}
 \caption{TESS light curve of NGC 6297. The red diamonds are raw 30-m FFI data and the silver diamonds are TESS data stacked in 1-day bins. The alternating shading represents the various sectors for which TESS observed the source. The vertical blue line is the time of first light obtained from our fit to the TESS data.}
 \label{fig:TESS_lc}
\end{figure}

We converted the measured fluxes into TESS-band magnitudes using an instrumental zero point of 20.44 electrons per second in the FFIs, based on the TESS Instrument Handbook \citep{vanderspek18}. The single TESS filter spans roughly $6000 - 10000$ \AA \ with an effective wavelength of $\sim7500$ \AA. We show the host-subtracted light curve for all TESS sectors in Figure~\ref{fig:TESS_lc}. We also show TESS photometry stacked in 1-day bins using a weighted average to compare to our ground-based photometry. Other than the ASASSN-20hx flare, there are no significant outbursts above a flux of $\sim0.1$ mJy in the FFIs. This corresponds to an AB mag of 18.9, placing strong constraints on optical host variability in the year prior to ASASSN-20hx. Indeed if we conduct the same exercise with the stacked photometry, the limit is even more constraining at  $\sim0.06$ mJy or an AB magnitude of 19.5. The root-mean-squared (RMS) values are 44 and 36 $\mu$Jy respectively for the FFIs and the stacked data. We can also use the stacked photometry to compute an approximate Eddington ratio. If we use the scatter of the stacked TESS photometry as a rough estimate of pre-existing AGN variability, we find a luminosity limit in the TESS band of $\lesssim 6 \times 10^{40}$ erg s$^{-1}$. This corresponds to an Eddington ratio of $\lesssim 6 \times 10^{-6}$ using the SMBH mass computed from galaxy scaling relations. Although this is a rough estimate, using only a single photometric band, it is clear that the host galaxy of ASASSN-20hx was not previously a highly variably AGN.

\subsection{Additional Ground-Based Photometry}

In addition to the ASAS-SN and TESS survey photometry, we obtained photometric follow-up observations from several ground-based observatories. We used the Las Cumbres Observatory Global Telescope (LCOGT) network \citep{brown13} 1-m and 2-m telescopes located at Haleakal\=a and McDonald Observatory for $BVgri$ observations, the Liverpool 2-m telescope at Observatorio del Roque de los Muchachos for $ugriz$ observations, the Wendelstein Observatory 2-m  Fraunhofer telescope for $ugizJH$ observations, and images from amateur astronomer Richard Post (RP) for $BVgri$ observations taken with a 24-in and 32-in telescope. After applying appropriate flat-field corrections, we obtained the astrometry for each image using astrometry.net \citep{barron08,lang10}.

For our ground-based data we used {\tt apphot} to measure 5\farcs{0} aperture magnitudes of the host plus transient emission, and subtracted the 5\farcs{0} host flux synthesised from our \textsc{FAST} fit in the appropriate filter to isolate the transient flux. Although there were archival SDSS images in the $ugriz$ filters, we found that host subtraction yielded significant artifacts and scatter in the subtracted images and magnitudes. For each filter, we used SDSS stars in the field to calibrate our photometry, using the corrections from \citet{lupton05} to calibrate the $B$ and $V$ band magnitudes with the $ugriz$ data. We found an offset between the ground-based and ASAS-SN $g$-band data, which we corrected by shifting the median of ground-based $g$-band to match the median ASAS-SN $g$-band mag.

We measured the centroid position of the transient in a host-subtracted LCOGT  $g$-band image taken near peak using the IRAF {\tt imcentroid} package. This yielded a position of $(\alpha,\delta)=($17:03:36.560$,+$62:01:32.18$)$. We used the archival SDSS $g$-band image to measure the position of the nucleus of NGC 6297, finding $(\alpha,\delta)=($17:03:36.555$,+$62:01:32.22$)$. This gives an angular offset of 0\farcs{094}$\pm$0\farcs{222}, where the uncertainty is due to uncertainty in the centroid positions of ASASSN-20hx and the host nucleus. As the transient position is consistent with the nucleus, we do not include the uncertainty due to the astrometric solution, which would only inflate the total error. At the host distance, this offset corresponds to a physical distance of $33.2\pm78.4$~pc.

\begin{deluxetable}{cccccc}
\tablewidth{240pt}
\tabletypesize{\footnotesize}
\tablecaption{Host-Subtracted Photometry of ASASSN-20hx}
\tablehead{
\colhead{MJD} &
\colhead{dMJD$_{hi}$} &
\colhead{dMJD$_{lo}$} & 
\colhead{Filter} &
\colhead{Magnitude} &
\colhead{Uncertainty} }
\startdata
59049.85 & 0.75 & 0.71 & $UVW2$ & 15.72 & 0.03 \\
59052.41 & 0.65 & 1.06 & $UVW2$ & 15.72 & 0.02 \\
59054.52 & 0.00 & 0.00 & $UVW2$ & 15.77 & 0.04 \\
\ldots & \ldots & \ldots & \ldots & \ldots & \ldots  \\
59126.00 & 0.00 & 0.00 & $z$ & 15.98 & --- \\
59137.89 & 0.00 & 0.00 & $z$ & 16.63 & 0.35 \\
59149.00 & 0.00 & 0.00 & $z$ & 16.26 & --- \\
\enddata 
\tablecomments{Host-subtracted magnitudes and 3$\sigma$ upper limits for all follow-up photometry. The MJD errors correspond to the range over which data were stacked. All magnitudes are corrected for Galactic extinction and presented in the AB system. Only a small section of the table is displayed here to show the format. We do not show the $J$- and $H$- band data as they are mostly upper limits and any detections are likely spurious. The full table can be found online as an ancillary file.} 
\label{tab:phot} 
\end{deluxetable}

\subsection{\textit{Swift} Observations}
Sixty-two total Neil Gehrels Swift Gamma-ray Burst Mission (\textit{Swift}; \citealt{gehrels04}) target-of-opportunity (ToO) observations were carried out between 2020 July 19 and 2021 April 15 (Swift target ID 13617, PIs: Holoien, Hinkle). Some of these epochs were obtained from Swift guest investigator program 1619122 (PI: Holoien). These observations used the UltraViolet and Optical Telescope (UVOT; \citealt{roming05}) and the X-ray Telescope (XRT; \citealt{burrows05}) to study the multi-wavelength properties of the ANT.

\subsubsection{UVOT Observations}
For a majority of the observation epochs, \swift{} observed ASASSN-20hx with all six UVOT filters \citep{poole08}: $V$ (5425.3 \AA), $B$ (4349.6 \AA), $U$ (3467.1 \AA), $UVW1$ (2580.8 \AA), $UVM2$ (2246.4 \AA), and $UVW2$ (2054.6 \AA). The wavelengths quoted here are the pivot wavelengths calculated by the SVO Filter Profile Service \citep{rodrigo12}. Each epoch of UVOT data includes 2 observations in each filter, which we combined into one image for each filter using the HEASoft {\tt uvotimsum} package. We then used the {\tt uvotsource} package to extract source counts using a 5\farcs{0} radius region centered on the position of the ANT and background counts using a source-free region with radius of $\sim$110\farcs{0}. We converted the UVOT count rates into fluxes and magnitudes using the most recent calibrations \citep{poole08, breeveld10}.

Because the UVOT uses unique $B$ and $V$ filters, we used publicly available color corrections\footnote{\url{https://heasarc.gsfc.nasa.gov/docs/heasarc/caldb/swift/docs/uvot/uvot_caldb_coltrans_02b.pdf}} to convert the UVOT $BV$ data to the Johnson-Cousins system. We then corrected the UVOT photometry for Galactic extinction and removed host contamination by subtracting the corresponding 5\farcs{0} host flux in each filter, as we did with the ground-based $BV$ data.

\begin{figure*}
\centering
 \includegraphics[width=0.99\textwidth]{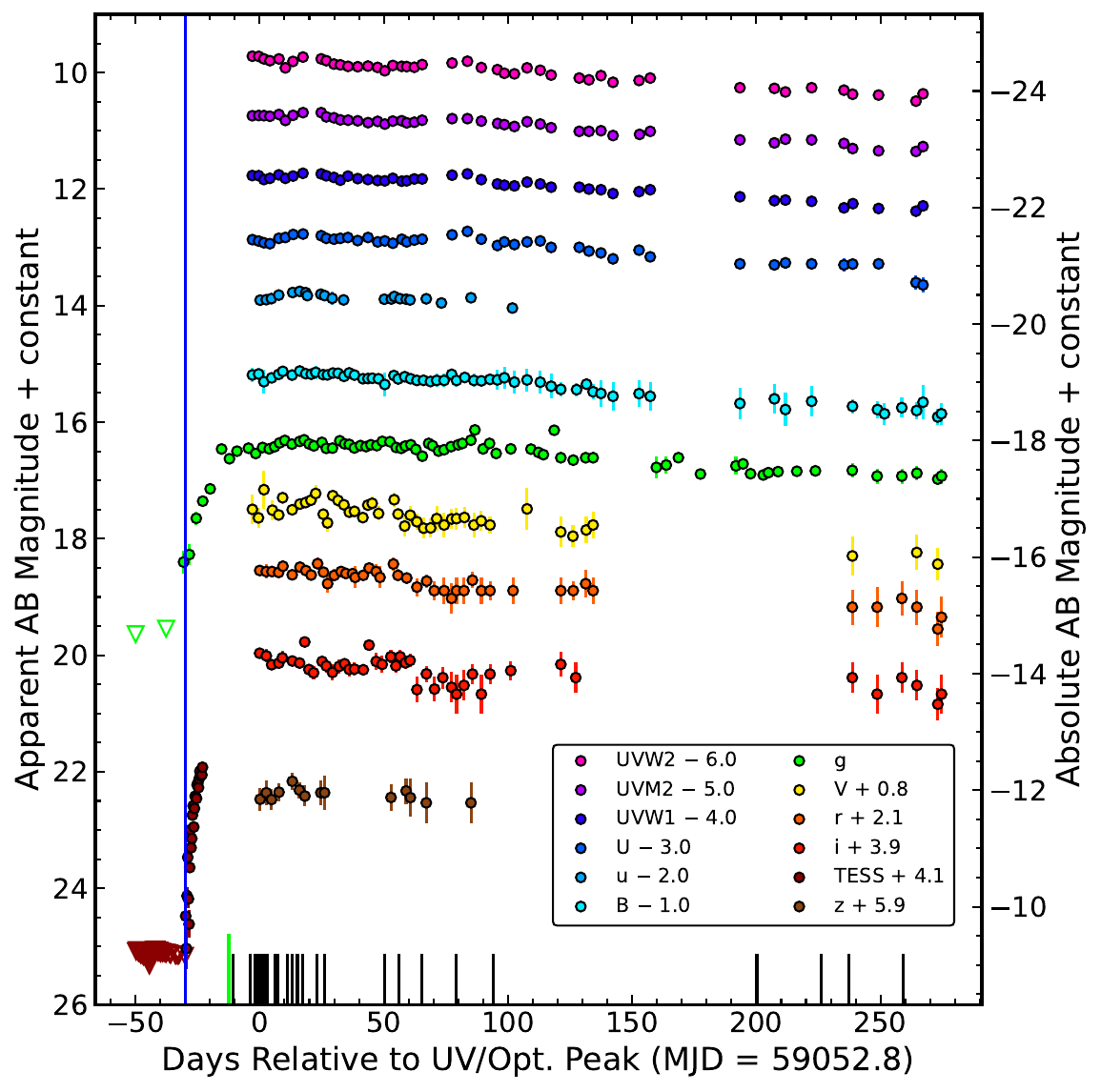}
 \caption{Host-subtracted UV and optical light curves of ASASSN-20hx, showing ASAS-SN ($g$), Swift (UV+$UBV$), Las Cumbres Observatory 1-m telescopes ($BVgri$), Liverpool Telescope ($ugriz$), RP ($BVgri$), and Wendelstein ($ugiz$) photometry stacked in 2-day bins. The TESS photometry is shown in 6-hour bins. The photometry spans from roughly 50 days prior to peak (MJD = 59052.8) to roughly 275 days after in observer-frame days. Horizontal error bars on the data indicate the date range of observations stacked to obtain deeper limits and higher S/N detections, although they are often smaller than the symbols. Open symbols indicate 3$\sigma$ upper limits and are only shown for the early-time data. The green bar on the x-axis marks the epoch of ASAS-SN discovery. The black bars along the x-axis show epochs of spectroscopic follow-up. The blue line is the estimated time of first light (see \S \ref{lc}) with uncertainty comparable to the line width. The time over which the first ASAS-SN detection is stacked is consistent with the time of first light. All data are corrected for Galactic extinction and shown in the AB system.}
 \label{fig:opt_lc}
\end{figure*}

Figure \ref{fig:opt_lc} shows the extinction-corrected, host-subtracted light curves of ASASSN-20hx. The photometry spans from the shortest $UVW2$ (2054.6 \AA) band of Swift to $z$ band ($\sim$ 8947 \AA) from LT and Wendelstein and includes data ranging from 30 days prior to peak to 275 days after peak. All the UV and optical photometry shown in Figure~\ref{fig:opt_lc}, and the limits not shown in this figure are presented in Table~\ref{tab:phot}.

\subsubsection{XRT Observations}
ASASSN-20hx was also observed using the \textit{Swift} X-ray Telescope in photon-counting mode. All observations were reduced using the standard filter and screening criteria specified in the \textit{Swift} XRT reduction guide\footnote{\url{http://swift.gsfc.nasa.gov/analysis/xrt_swguide_v1_2.pdf}} and the most up to date calibration files. Here, we reprocessed level one XRT data using the task \textsc{xrtpipeline} version 0.13.2, producing cleaned event files and exposure maps for all observations.  

To extract background-subtracted count rates from each individual observation, we used a 20\farcs{0} source region centered on the position of ASASSN-20hx and a source-free background region with a radius of 150\farcs{0} centered at ($\alpha$,$\delta$)=($17^{h}02^{m}36.19^{s},+62^{\circ}02'22.64''$). All extracted count rates were corrected for the encircled energy fraction\footnote{A 20\farcs{0} source radius contains $\sim80\%$ of the source counts at 1.5\,keV, assuming an on-axis pointing \citep{moretti04}}. 

To increase the S/N of our observations, we also combined our individual \textit{Swift} XRT observations using \textsc{XSELECT} version 2.4k. Here we combined the observations into 12 time bins spanning the $\sim$300 day \textit{Swift} observing campaign. This allowed us to extract spectra with $\sim500$ background subtracted counts using the task \textsc{xrtproducts} version 0.4.2 and the source and background regions defined above. Ancillary response files (ARF) were generated using the task \textsc{xrtmkarf} and the individual exposure maps that were generated by \textsc{xrtpipeline} and then merged using \textsc{ximage} version 4.5.1. The
response matrix files (RMFs) are ready-made files which we obtained from the most recent calibration database. Each spectrum was grouped to have a minimum of 15 counts per energy bin using the \textsc{ftools} command \textsc{grppha}.

\subsection{\textit{NICER} Observations}
ASASSN-20hx was also observed using the X-ray timing instrument (XTI) onboard the Neutron star Interior Composition ExploreR (NICER: \citealt{gendreau12}), which is a external payload on the International Space Station. NICER offers high spectral ($\sim85$ eV at 1 keV) and time resolution ($\sim100$ ns) observations in the 0.2-12 keV energy range and has been used to observe a number of nuclear transients \citep{trakhtenbrot19b, 2020ATel14248....1P,  2020ATel14036....1P, 2020ATel13864....1P, hinkle21a, payne21b, ricci21, 2021MNRAS.504..792C, 2021ApJ...912..151W}. ASASSN-20hx was observed a total of 41 times between 2020 July 25 and 2020 November 01 (ObsIDs: 3573010101$-$3573014102, PI: Auchettl), for a total cumulative exposure of 67.6 ks. 

The data were reprocessed using the \textsc{nicerdas} version 7a and the task \textsc{nicerl2}. Here standard filtering criteria were used\footnote{See \citet{2019ApJ...887L..25B} or Section 2.7 of \citet{hinkle21b}}, as well as the latest gain and calibrations files. Time averaged spectra and count rates were extracted using \textsc{xselect}, while we took advantage of ready made ARF (nixtiaveonaxis20170601v004.arf) and RMF (nixtiref20170601v002.rmf) files that are available with the \textit{NICER} CALDB. All spectra was grouped using a minimum of 20 counts per energy bin. As \textit{NICER} is a non-imaging instrument, background spectra were generated using the background modeling tool \textsc{nibackgen3C50}\footnote{\url{https://heasarc.gsfc.nasa.gov/docs/nicer/tools/nicer_bkg_est_tools.html}}.

To analyse the spectral data from both \textit{Swift} and \textit{NICER}, we used the X-ray spectral fitting package (XSPEC) version 12.11.1 \citep{arnaud96} and $\chi^{2}$ statistics. Both the \textit{Swift} and \textit{NICER} data and their analysis are further discussed in \S \ref{sec:xray}.

\begin{deluxetable}{cccccc}
\tablewidth{240pt}
\tabletypesize{\footnotesize}
\tablecaption{X-ray Luminosities and Hardness Ratios of ASASSN-20hx}
\tablehead{
\colhead{MJD} &
\colhead{log[L (erg s$^{-1}$)]} &
\colhead{L Error} & 
\colhead{HR} &
\colhead{HR Error} &
\colhead{Satellite} }
\startdata
59049.14 & 42.32 & 0.04 & -0.73 & 0.13 & \swift \\
59050.60 & 42.30 & 0.04 & -0.60 & 0.07 & \swift \\
59051.35 & 42.06 & 0.05 & -0.62 & 0.11 & \swift \\
\ldots & \ldots & \ldots & \ldots & \ldots & \ldots \\
59317.09 & 41.74 & 0.10 & -0.69 & 0.31 & \swift \\
59319.93 & 41.60 & 0.15 & -0.79 & 0.00 & \swift \\
59346.30 & 41.96 & 0.05 & -0.43 & 0.05 & \swift \\
\enddata 
\tablecomments{X-ray luminositites and hardness ratios with associated uncertainties. The hardness ratio is defined as (H$-$S)/(H+S), where we define hard counts H as the number of counts in the 2-10 keV range and soft counts S are the number of counts in the 0.3-2 keV. The last column reports the source of the data for each epoch. Only a small section of the table is displayed here. The full table can be found online as an ancillary file.} 
\label{tab:xray_lum_hr} 
\end{deluxetable}

\subsection{Spectroscopic Observations}

In addition to our classification spectra obtained from SPRAT and LRIS, we obtained follow-up spectra of ASASSN-20hx with the SuperNova Integral Field Spectrograph \citep[SNIFS;][]{lantz04} on the 88-inch University of Hawaii telescope (UH88), LRIS on the 10-m Keck I telescope, SPRAT on the 2-m Liverpool Telescope, and the Multi-Object Double Spectrographs \citep[MODS;][]{pogge10} on the Large Binocular Telescope \citep[LBT; ][]{hill06}. Three of our spectra were obtained prior to peak light and nineteen were obtained after peak. Most of the spectra were reduced and calibrated with standard IRAF procedures, such as bias subtraction, flat-fielding, 1-D spectroscopic extraction, and wavelength calibration. The LRIS spectra were reduced in part using LPipe \citep{perley19}. The SNIFS spectra were calibrated using a custom reduction pipeline. The flux calibration for our observations was initially done using standard star spectra obtained on the same nights as the science spectra and then refined using our follow-up photometry and blackbody fit results.

\begin{figure*}
\centering
 \includegraphics[width=0.98\textwidth]{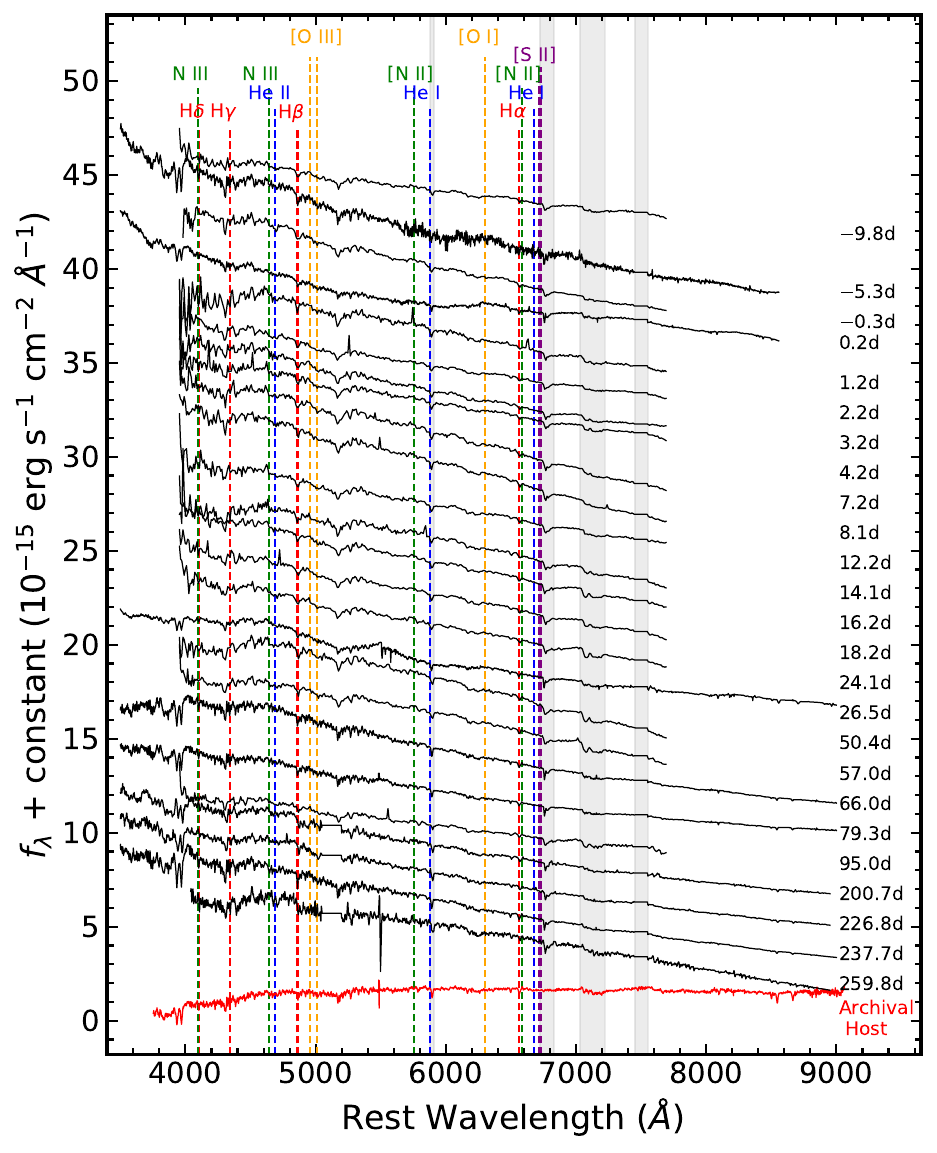}
 \caption{Optical spectroscopic evolution of ASASSN-20hx spanning from 10 days prior to peak UV/optical emission (top) until 259 days after peak (bottom). These spectra are calibrated using the photometry presented in Figure \ref{fig:opt_lc} as well as the corresponding blackbody fits. The archival host spectrum from SDSS is shown in red at the very bottom. The vertical gray bands mark atmospheric telluric features. The strong telluric feature between $\sim$ 7400 - 7550 \AA \ and several artifacts in individual spectra been masked. The vertical lines mark spectral features common in TDEs and AGNs, with hydrogen lines in red, helium lines in blue, nitrogen lines in green, sulfur lines in purple, and oxygen lines in orange.}
 \label{fig:opt_spec}
\end{figure*}

All the classification and follow-up optical spectra for ASASSN-20hx are presented in Figure \ref{fig:opt_spec}. From top to bottom, the optical spectrum remains blue and largely featureless throughout the evolution. Nonetheless, the spectra do show some absorption lines of H$\alpha$, H$\beta$, \ion{Ca}{2} H and K, and \ion{Na}{1} D, although these are likely contamination from the bright host galaxy. The locations of several emission lines commonly seen in TDEs and AGNs are marked with vertical dashed lines. None of these lines appear in emission throughout the time period probed by these spectra.

In addition to our optical spectra, we obtain three near-infrared (NIR) spectra of ASASSN-20hx with SpeX \citep{rayner03} on the NASA Infrared Telescope Facility (IRTF). These data were obtained in the short cross-dispersed mode giving us coverage over roughly the $zYJHK$ bands at moderate resolution. These spectra were reduced and calibrated with standard IRAF procedures, such as bias subtraction, flat-fielding, 1-D spectroscopic extraction, and wavelength calibration. When possible, a telluric standard from the same night was used to remove atmospheric contamination. 

\subsection{Sub-mm Observations}

We obtained sub-mm observations in the 850 $\mu$m-band (353 GHz) using the Submillimetre Common-User Bolometer Array 2 \citep[SCUBA-2;][]{holland13} on the James Clerk Maxwell Telescope (JCMT). As the expected emission from ASASSN-20hx would be a point source, we used a simple constant velocity Daisy mapping pattern.  These data were calibrated and reduced using the default SCUBA-2 pipeline at the Canadian Astronomical Data Centre\footnote{\url{https://www.cadc-ccda.hia-iha.nrc-cnrc.gc.ca/en/}} (CADC) with phase and flux calibrators taken on the same night as the source observations. We discuss the derived sub-mm limits of ASASSN-20hx relative to known TDEs in Section \ref{sec:radio_analysis}.

\section{Analysis}\label{analysis}

Here we present analysis of the UV/optical light curve, sub-mm limits, optical and NIR spectroscopy, the spectral energy distribution, and X-rays to place ASASSN-20hx in context with other transients. While the long plateau phase of the light curve is reminiscent of Type IIP supernovae, the length of this phase is longer than even the most extreme of these SNe \citep[e.g.,][]{sanders15, reguitti21}. Additionally, the spectra are inconsistent with a SN II as they lack hydrogen features. The spectra are formally consistent with a Type Ic supernova, but the strong, persistent X-ray emission and the long-lived light curve are not \citep[e.g.,][]{taddia18}. Additionally, ASASSN-20hx bears some similarities to the constant temperature and lack of strong emission lines in some luminous red novae \citep[e.g.,][]{blagordnova20}. However, this scenario is not likely due to ASASSN-20hx being much hotter and several orders of magnitude more luminous. Because the observed properties of ASASSN-20hx are incompatible with known supernovae and stellar transients, we focus the remainder of this manuscript on distinguishing between TDE and AGN activity. 

\subsection{Light Curve}\label{lc}

Using Markov Chain Monte Carlo (MCMC) methods, we fit each of the epochs with \swift UV photometry as a blackbody to obtain the bolometric luminosity, temperature, and effective radius of ASASSN-20hx. To keep our fits relatively unconstrained, we ran each of our blackbody fits with flat temperature priors of 10000 K $\leq$ T $\leq$ 55000. We find that a blackbody model adequately describes the UV/optical emission, with a median reduced $\chi^2$ of $\sim 1.3$, similar to or better than previous TDE candidates \citep[e.g.,][]{holoien14b}. To obtain the time of peak UV/optical luminosity, we fit a parabola to the light curve created by bolometrically correcting the ASAS-SN $g$-band light curve using these blackbody fits, excluding upper limits. Because the curve is quite flat near peak, we fit the parabola in a narrow range between MJD = 59029.9 and MJD = 59061.0. Despite the flatness of the light curve, we find a median reduced $\chi^2$ of 1.4, indicating that the parabola is an acceptable fit. We generated 10,000 realizations of the bolometric light curve in this date range with each luminosity perturbed by its uncertainty assuming Gaussian errors. We then fit a parabola to each perturbed light curve and took the median value as the peak and 16th and 84th percentiles as the uncertainties in peak time. Using this procedure, we find the time of peak bolometric luminosity to be MJD $= 59052.8^{+0.8}_{-0.6}$. If we instead fit just the $g$-band data, we find a peak time of MJD $= 59065.9^{+0.9}_{-0.8}$, suggesting that like some TDEs \citep{holoien18a, holoien19a, hinkle21a} ASASSN-20hx may have peaked earlier in bluer bands, although we do not have early enough \swift coverage to test this directly.

\begin{figure}
\centering
 \includegraphics[width=.48\textwidth]{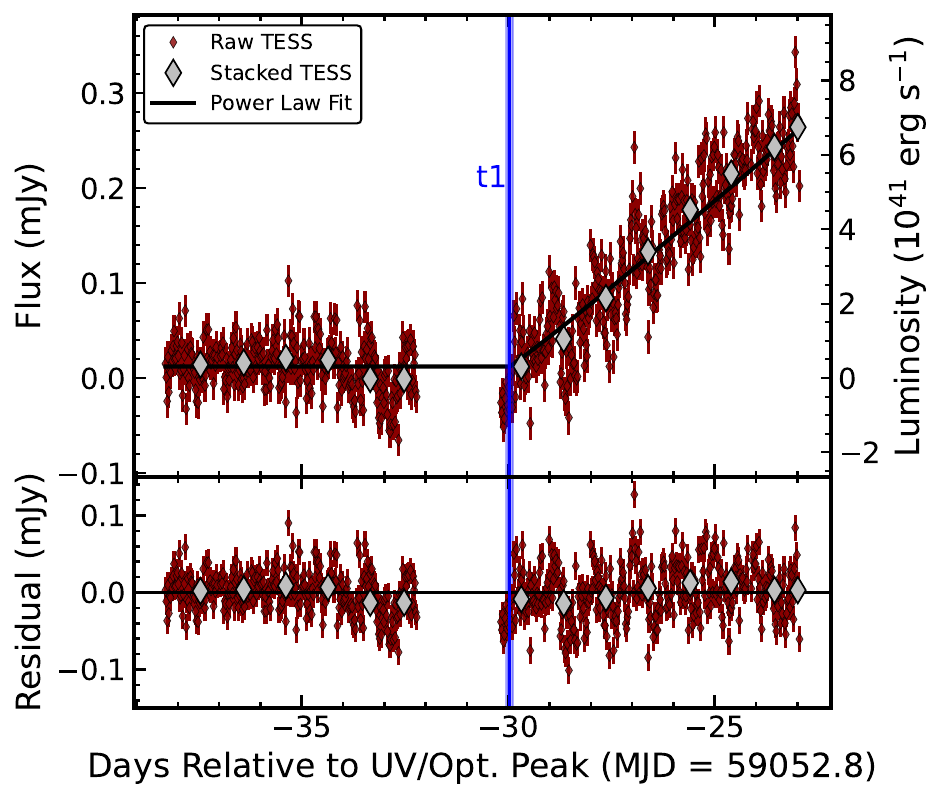}\hfill
 \caption{\textit{Top panel}: Stacked (silver) and raw (red) TESS light curve and best-fit power-law model in black. This power-law fit yields a time of first light of $t_1$ = $59022.8 \pm 0.1$ and a power-law index of $\alpha = 1.05 \pm 0.06$. The blue line shows the fit time of first light with the uncertainty comparable to the line width. \textit{Bottom panel}: Residuals between the data and best-fit power-law model.}
 \label{fig:rise}
\end{figure}

ASASSN-20hx is the first ANT for which the early-time rise was caught in an active TESS sector. We fit the early-time rise as a power law with

\begin{equation}
f = z \text{ for $t < t_1$, and}
\end{equation}

\begin{equation}
f = z + h\bigg(\frac{t - t_1}{\text{observed days}}\bigg)^\alpha \text{for $t > t_1$}
\end{equation}

\noindent This model fits for the zero point $z$, the time of first-light $t_1$, a flux scale $h$, and the power-law index $\alpha$. An MCMC fit yields best-fit parameters of $z = 12.1 \pm 1.0 \  \mu\text{Jy}, \ h = 32.5^{+3.5}_{-3.2} \ \mu\text{Jy}, \ t_1 \text{(MJD)} = 59022.8 \pm 0.1, \ \text{and} \ \alpha = 1.05 \pm 0.06$ and are shown in Figure \ref{fig:rise}. While the best-fit time of first light appears to be very close to the downlink gap, we re-ran the analysis with the first 6 hours of post-gap TESS data removed and found parameters consistent with the original fit. From Figure \ref{fig:rise}, we see that the light curve rises from the time of first light to the peak UV/optical bolometric luminosity in $\sim30$ days, shorter than the rise to peak time measured for ASASSN-19bt \citep{holoien19c}, and the limits on the rise times for PS18kh \citep{holoien19b} and ASASSN-18pg \citep{holoien20}, but similar to the rise time for the TDE ASASSN-19dj \citep{hinkle21a}. From the best-fit time of first light, we find that ASAS-SN discovered this transient just over two weeks after the beginning of the flare.

The best-fit power-law index of $\alpha = 1.05 \pm 0.06$ for the rise of ASASSN-20hx is consistent with the rise slope for the recurring partial TDE ASASSN-14ko although the rise time for ASASSN-14ko is much shorter at $\sim 5$ days \citep{payne21}. Conversely, the rise slopes of the TDEs ASASSN-19bt \citep{holoien19b}, ASASSN-19dj \citep{hinkle21a}, and ZTF19abzrhgq \citep[AT2019qiz, ][]{nicholl20} are each consistent with a quadratic rise ($\alpha = 2)$.

Under the assumption that ASASSN-20hx might be a TDE, we used the Modular Open-Source Fitter for Transients \citep[\texttt{MOSFiT};][]{guillochon17b, mockler19} to fit the host-subtracted light curves of ASASSN-20hx to estimate physical parameters of the star, SMBH, and the interaction. \texttt{MOSFiT} uses models with multiple physical parameters to generate bolometric light curves of a TDE, and computes single-filter light curves from the bolometric light curves that are then fit to the observed data. \texttt{MOSFiT} then finds the highest likelihood match of the various parameters for a given model using one of various sampling methods. We ran the \texttt{MOSFiT} TDE model in the nested sampling mode to account for the large number of observations in several photometric filters.

\begin{figure*}
\begin{minipage}{\textwidth}
\centering
{\includegraphics[width=0.95\textwidth]{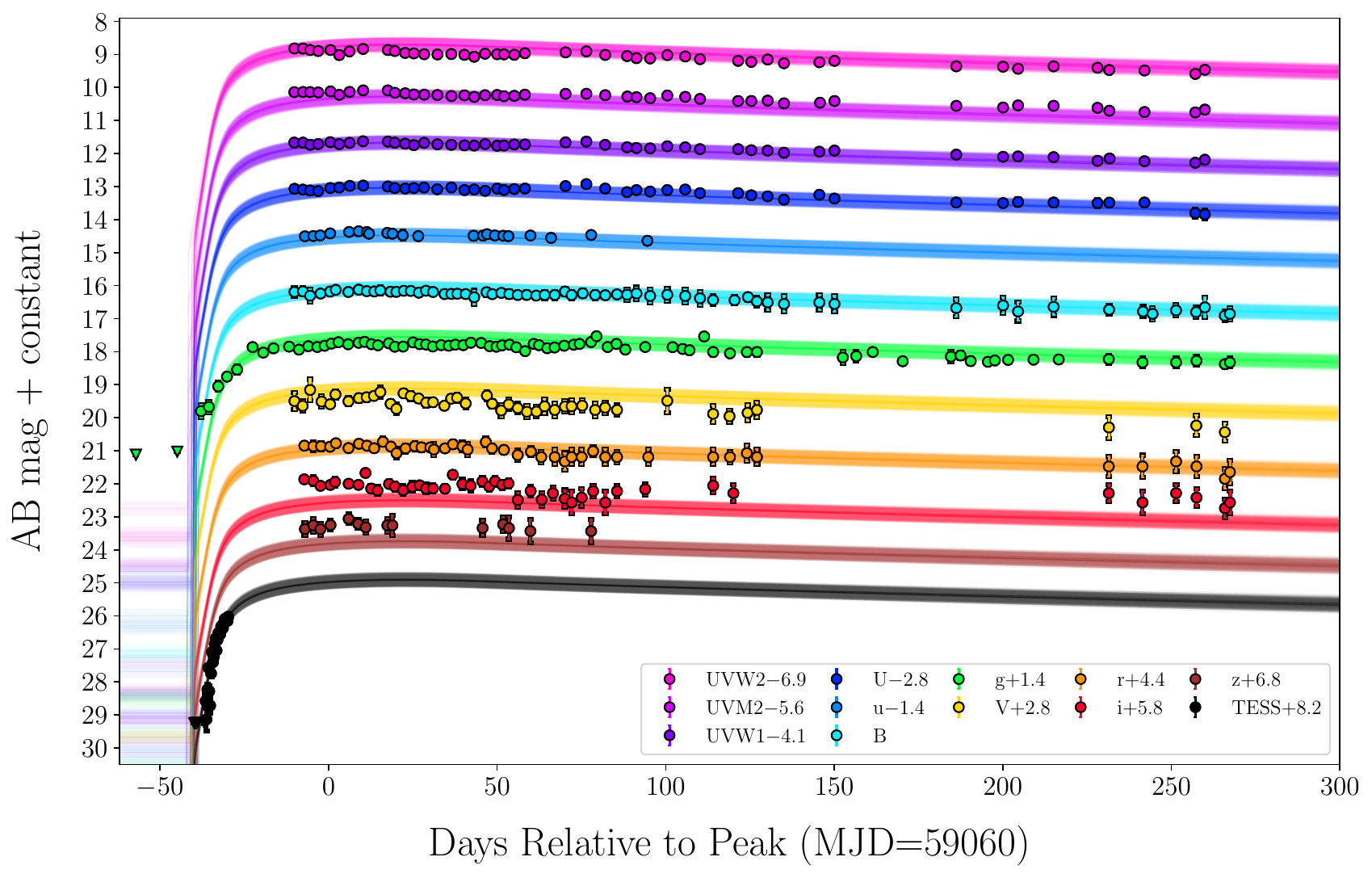}}
\caption{\texttt{MOSFiT} light curve fits and host-subtracted light curves. The $1-99$\% range of fits for each filter are shown as shaded regions with the median fit shown as a solid line. All detections are plotted as circles with $3\sigma$ upper limits plotted as downward triangles. The colors match those of Figure~\ref{fig:opt_lc}.}
\label{fig:mosfit_lc}
\end{minipage}
\end{figure*}

During initial testing of the \texttt{MOSFiT} fits to our dataset, we found that \texttt{MOSFiT} preferred a black hole mass of $M_h\sim10^6$~\msun. As this is nearly 2 orders of magnitude below the mass we estimated from the stellar mass of the host galaxy, we performed our final \texttt{MOSFiT} fits using a prior of $7\leq log(M_h) \leq8.5$ on the mass of the SMBH. Figure \ref{fig:mosfit_lc} shows the \texttt{MOSFiT} multi-band fits to the light curves of ASASSN-20hx.

\texttt{MOSFiT} is one of the only available tools for fitting TDE emission in detail, and appears to reproduce events like ASASSN-20hx, with relatively smooth light curves well. \texttt{MOSFiT} does a reasonable job of fitting the data throughout the event, both near peak and at late times. Given the rapid rise of ASASSN-20hx relative to its long decay, the rise is not particularly well constrained by the \texttt{MOSFiT} model. The strong agreement between the late-time data and the model may suggest that if ASASSN-20hx is a TDE it may be predominantly accretion-powered.

\begin{deluxetable}{lrc}
\tablewidth{240pt}
\tabletypesize{\footnotesize}
\tablecaption{\texttt{MOSFiT} TDE Model Parameter Fits}
\tablehead{
\colhead{Quantity} &
\colhead{Value} &
\colhead{Units} }
\startdata
$\log{R_{\textrm{ph0}}}$ (photosphere power-law constant) & $-0.74^{+0.40}_{-0.40}$& --- \\
$\log{T_{\textrm{viscous}}}$ (viscious delay timescale) & $-1.49^{+1.00}_{-1.00}$ & days \\
$b$ (scaled impact parameter $\beta$) & $0.81^{+0.89}_{-0.81}$ & --- \\
$\log{M_{h}}$ (SMBH mass) & $7.03^{+0.20}_{-0.20}$& \msun \\
$\log{\epsilon}$ (efficiency) & $-0.48^{+0.68}_{-0.68}$ & --- \\
$l$ (photosphere power-law exponent) & $0.30^{+0.21}_{-0.21}$ & --- \\
$\log{n_{\textrm{H,host}}}$ (local hydrogen column density) & $21.07^{+0.01}_{-0.01}$ & cm$^{-2}$ \\
$M_\star$ (stellar mass) & $0.36^{+3.89}_{-0.33}$ & \msun \\
$t_\textrm{exp}$ (time of disruption) & $-1.67^{+15.00}_{-15.02}$ & days \\
$\log{\sigma}$ (model variance) & $-0.67^{+0.02}_{-0.02}$ & --- \\
\enddata 
\tablecomments{Best-fit values and $1-99$\% ranges for the \texttt{MOSFiT} TDE model parameters. Units are listed where appropriate. The listed uncertainties include both statistical uncertainties from the fit and the systematic uncertainties listed in Table 3 of \citet{mockler19}. The systematic error on $b$ of 0.88, is larger than our estimated value.}
\label{tab:mosfit_params} 
\end{deluxetable}

Table~\ref{tab:mosfit_params} shows the median and $1-99$\% range for the \texttt{MOSFiT} TDE model parameters. The model parameters are generally very well constrained, with statistical uncertainties from the fit being significantly smaller than the systematic uncertainties of the model \citep[see Table 3 of][]{mockler19}. The black hole mass and stellar mass given by \texttt{MOSFiT} are log$(M_h / \msun)=7.03 \pm 0.20$ and $M_\star=0.36^{+3.89}_{-0.33}$~\msun, respectively. This black hole mass is smaller than the value estimated from scaling relations using our host stellar mass and is close to the lower bound of our prior constraints on the mass, which is perhaps unsurprising. The mass of the star is similar to the masses obtained for the TDEs modelled by \citet{mockler19}. As expected from the long overall timescale for ASASSN-20hx and the relatively slow rise to peak, \texttt{MOSFiT} prefers a fit in which the star was only partially disrupted in the encounter, with $b = 0.81$. However, for an interaction with a low-mass star and a $b$ value less than 1, the systematic uncertainties on $b$ are quite significant, and the model is consistent with both full disruption ($b<1$) and minimal partial disruption ($b<<1$) scenarios.

We also performed fits without using our prior on the SMBH mass, and we find that the resulting light curve fits are comparable to those from our constrained fit that are shown in Figure~\ref{fig:mosfit_lc}. Given that the fit is not clearly better when the SMBH mass is unconstrained, we determined that the results of our constrained fit are valid, and present those in full above. However, for the sake of clarity, we note that in the unconstrained fits the best-fit masses and $b$ values are log$(M_h / \msun)=6.23^{+0.23}_{-0.22}$, $M_\star=0.50^{+1.84}_{-0.39}$~\msun, and $b=1.03^{+0.22}_{-0.88}$.

\texttt{MOSFiT} assumes that the observed TDE is predominantly tied to rapid circularization and accretion onto the SMBH. If we instead want to understand the emission of ASASSN-20hx near peak in the context of a TDE powered by stream-stream collisions, we used \texttt{TDEmass} \citep{ryu20}. Unlike \texttt{MOSFiT}, \texttt{TDEmass} assumes that the UV/optical emission is shock-powered and extracts the SMBH and stellar mass based on the observed peak luminosity and temperature at peak. Using our peak luminosity of $(3.15 \pm 0.04) \times 10^{43}$ erg s$^{-1}$ and temperature at peak of $22,800^{+910}_{-875}$ K, we obtain a SMBH mass of $9.1^{-1.0}_{+2.5} \times 10^5$ \msun and a disrupted stellar mass of $0.62_{-0.01}^{+0.05}$ \msun. Given the stellar mass of the host galaxy, this SMBH mass is several orders of magnitude lower than expected. The disrupted stellar mass is consistent with other TDEs \citep{ryu20} and consistent with the \texttt{MOSFiT} result.

The estimated SMBH masses from both \texttt{MOSFiT} and \texttt{TDEmass} are lower than the mass estimated from the host galaxy mass and scaling relations. This may suggest that a TDE origin for ASASSN-20hx is disfavored. However, for the case of the \texttt{MOSFiT} model with a prior on mass, we find that when incorporating uncertainties both on the scaling relations and the \texttt{MOSFiT} result that the SMBH masses are consistent to within $\sim2\sigma$.

\subsection{Radio Constraints} \label{sec:radio_analysis}
\begin{figure}
\centering
 \includegraphics[width=.48\textwidth]{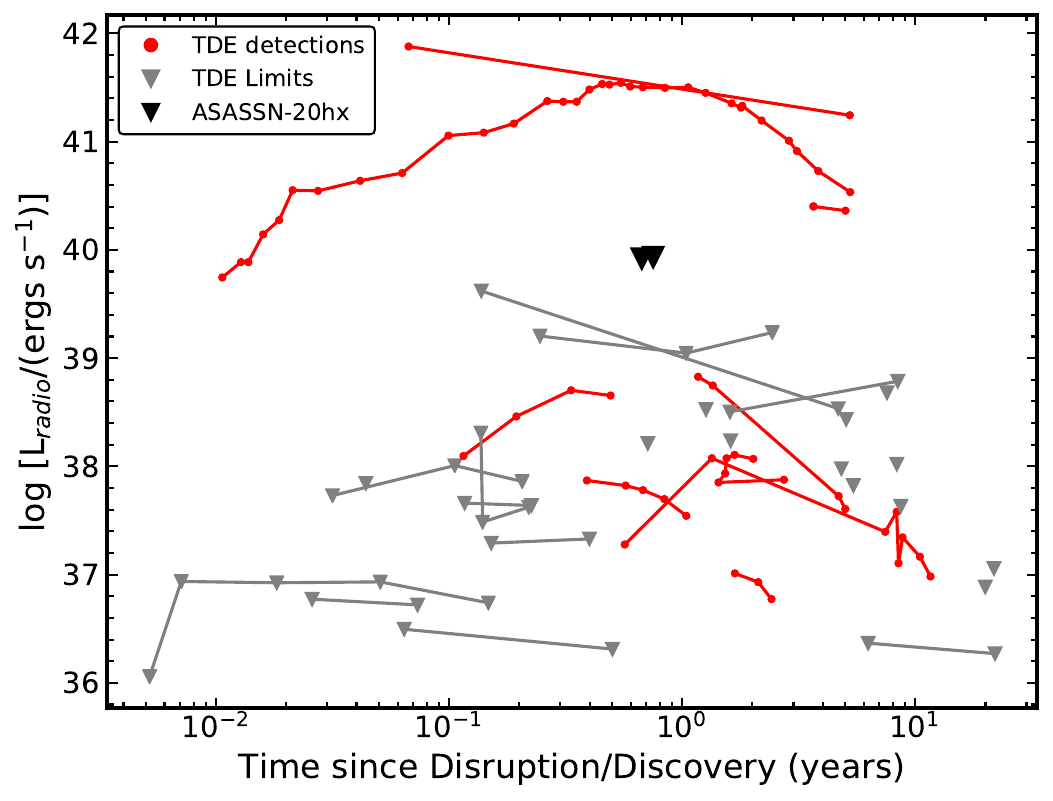}\hfill
 \caption{$\sim5$ GHz radio light curves of TDEs from \citet{alexander20}. The red lines show detected radio emission from TDEs and the gray lines show upper limits. The two black downward facing trangles are the 3$\sigma$ upper limits for the radio emission of ASASSN-20hx.}
 \label{fig:radio_lc}
\end{figure}

In addition to the optical light curve of ASASSN-20hx, we can compare the sub-mm limit we obtained with JCMT to radio observations of TDEs. At phases of $\sim245$ and $\sim274$ days from first light, ASASSN-20hx was not detected in the sub-mm, with 3$\sigma$ upper-limits of 8.9 mJy and 8.5 mJy respectively. As most observations of TDEs occur in the radio rather than the sub-mm, we converted our 353 GHz flux to a 5 GHz flux for more direct comparison \citep[e.g.,][]{alexander20}. To do this, we used the radio light curve of ZTF19aapreis from \citet{cendes21} as an approximate radio light curve model for a normal TDE. We adopted the basic assumptions of $\nu_m = 0.1$ GHz and $p = 2.7$ from \citet{cendes21}. For each epoch in their radio light curve, we used their measured $\nu_a$ and peak flux values to scale the expected radio/sub-mm SED. We then measured the flux ratio between the 5 GHz and 353 GHz at each epoch. Finally, we interpolated this scaling to find the factor needed to estimate at 5 GHz flux at the phase of our JCMT epoch. This yielded a conversion factor of $f_{353 GHz} / f_{5 GHz} = 0.0343$ for the JCMT epoch which occurred $\sim245$ days after first light and $f_{353 GHz} / f_{5 GHz} = 0.0317$ for $\sim274$ days after first light. Therefore, given our 353 GHz upper-limit, we estimate limits of 259 mJy and 269 mJy at 5 GHz at the two epochs. At the distance of ASASSN-20hx, these correspond to a 5 GHz luminosity of $8.2 \times 10^{39}$ erg s$^{-1}$ and $8.5 \times 10^{39}$ erg s$^{-1}$. These estimates are shown in Figure \ref{fig:radio_lc}. While this estimate requires several assumptions about the type of transient and radio evolution, it is sufficient to show that ASASSN-20hx is considerably fainter than the jetted/relativistic TDEs Swift J2058+05 and Swift J1644+57. This likely suggests that ASASSN-20hx either did not launch a strong jet or that the jet was off-axis \citep{irwin15, yuan16}.

\subsection{Spectra}  \label{sec:spectra}

The spectra of ASASSN-20hx exhibit a blue continuum and are devoid of any emission line features. Indeed, even after subtracting the SDSS host spectrum from the various follow-up spectra, emission lines remain conspicuously absent. This type of evolution is unusual for both TDEs and AGN. TDEs commonly exhibit broad H emission \citep[e.g.,][]{holoien14b, holoien20, hinkle21a, hung20} and many TDEs also show broad He lines \citep[e.g.,][]{blagorodnova17, wevers19, hinkle21a} and/or metal lines of O and N \citep{leloudas19,vanvelzen21}. None of these lines are present in the spectra of ASASSN-20hx at any point in the evolution. Conversely AGNs tend to have H lines that are somewhat narrower than those seen in TDEs at a given SMBH mass and luminosity along with narrow forbidden lines of O, N, and S \citep{baldwin81, veilleux87}. 

\begin{figure*}
\centering
 \includegraphics[width=1.0\textwidth]{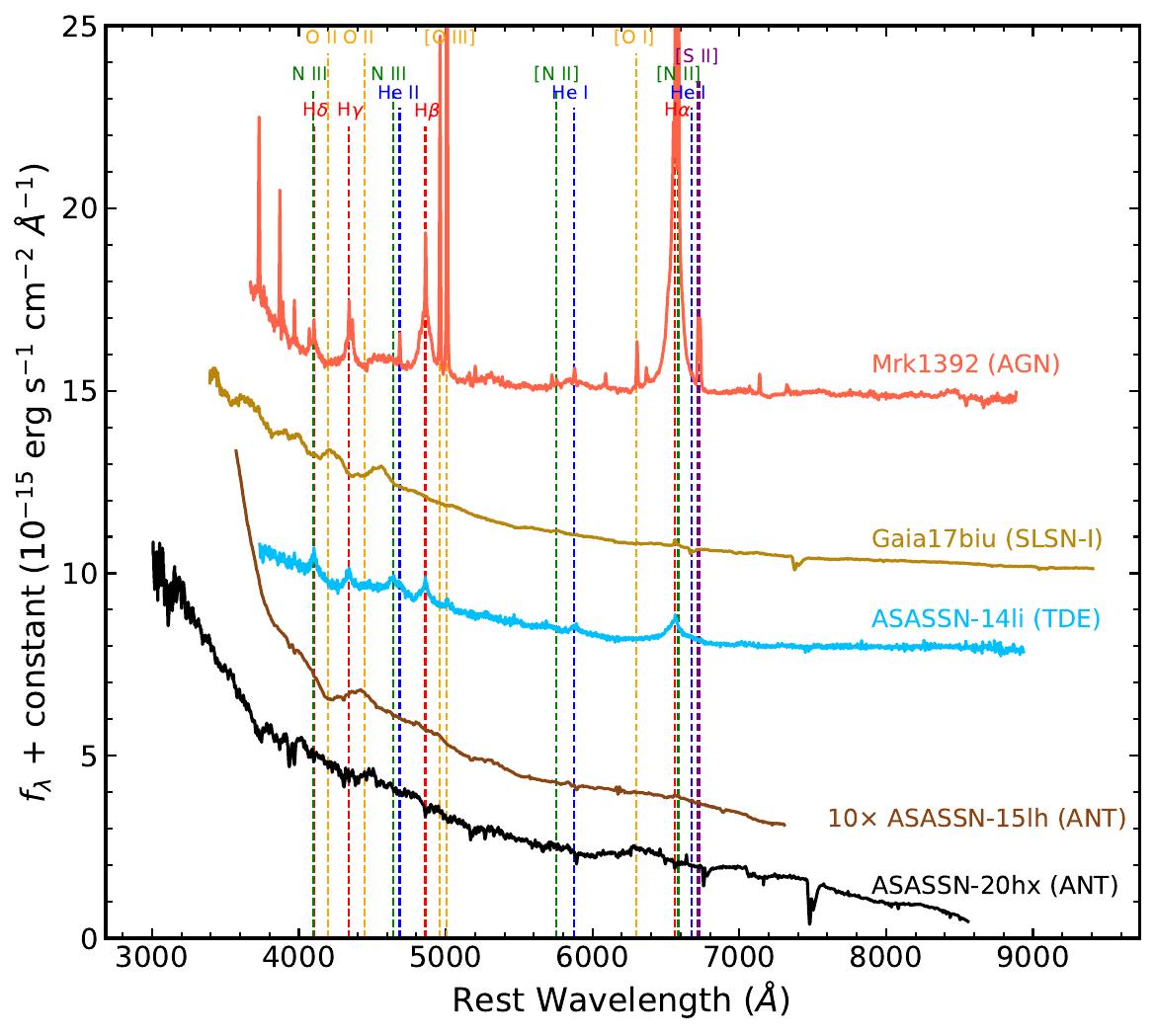}\hfill
 \caption{Spectrum of ASASSN-20hx as compared to the ANT ASASSN-15lh \citep[brown line;][]{dong16, leloudas16}, the TDE ASASSN-14li \citep[blue line;][]{holoien16a}, the SLSN-I Gaia17biu \citep[golden line;][]{bose18c}, and the Seyfert 1 Mrk1392 \citep[red line;][]{adelmanmccarthy06}. The spectra of the transients are all selected to be near peak emission. The vertical lines mark spectral features common in TDEs, AGNs, and SLSNe with hydrogen lines in red, helium lines in blue, nitrogen lines in green, sulfur lines in purple, and oxygen lines in orange.}
 \label{fig:comp_spec}
\end{figure*}

Figure \ref{fig:comp_spec} shows a spectrum of ASASSN-20hx near peak as compared to a well-studied TDE and a strong AGN. ASASSN-20hx shows none of the emission or absorption lines seen in these other classes of objects. Figure \ref{fig:comp_spec} also shows a spectrum of an SLSNe-I, which are largely featureless near peak except for \ion{O}{2} absorption \citep[e.g,][]{gal-yam12, bose18c, lunnan20}. This SLSN-I spectrum does not match the largely featureless nature of ASASSN-20hx, where no broad line features are present. Additionally, we compare ASASSN-20hx to the ANT ASASSN-15lh \citep[e.g.,][]{dong16, leloudas16}. While significantly more luminous, ASASSN-15lh has a similarly featureless spectrum. In Figure \ref{fig:comp_spec} we see that ASASSN-20hx has even fewer features than ASASSN-15lh, lacking the broad features near $\sim 4300$ \AA. Even if ASASSN-20hx is an example of a new AGN rapidly turning on, the lack of hydrogen emission would be unusual \citep[e.g.,][]{trakhtenbrot19a}. The only obvious case in which an AGN would not show emission lines is a blazar \citep[e.g.][]{paggi14}, although this is likely ruled out by the lack of detected sub-mm emission.

\begin{figure*}
\centering
 \includegraphics[width=1.0\textwidth]{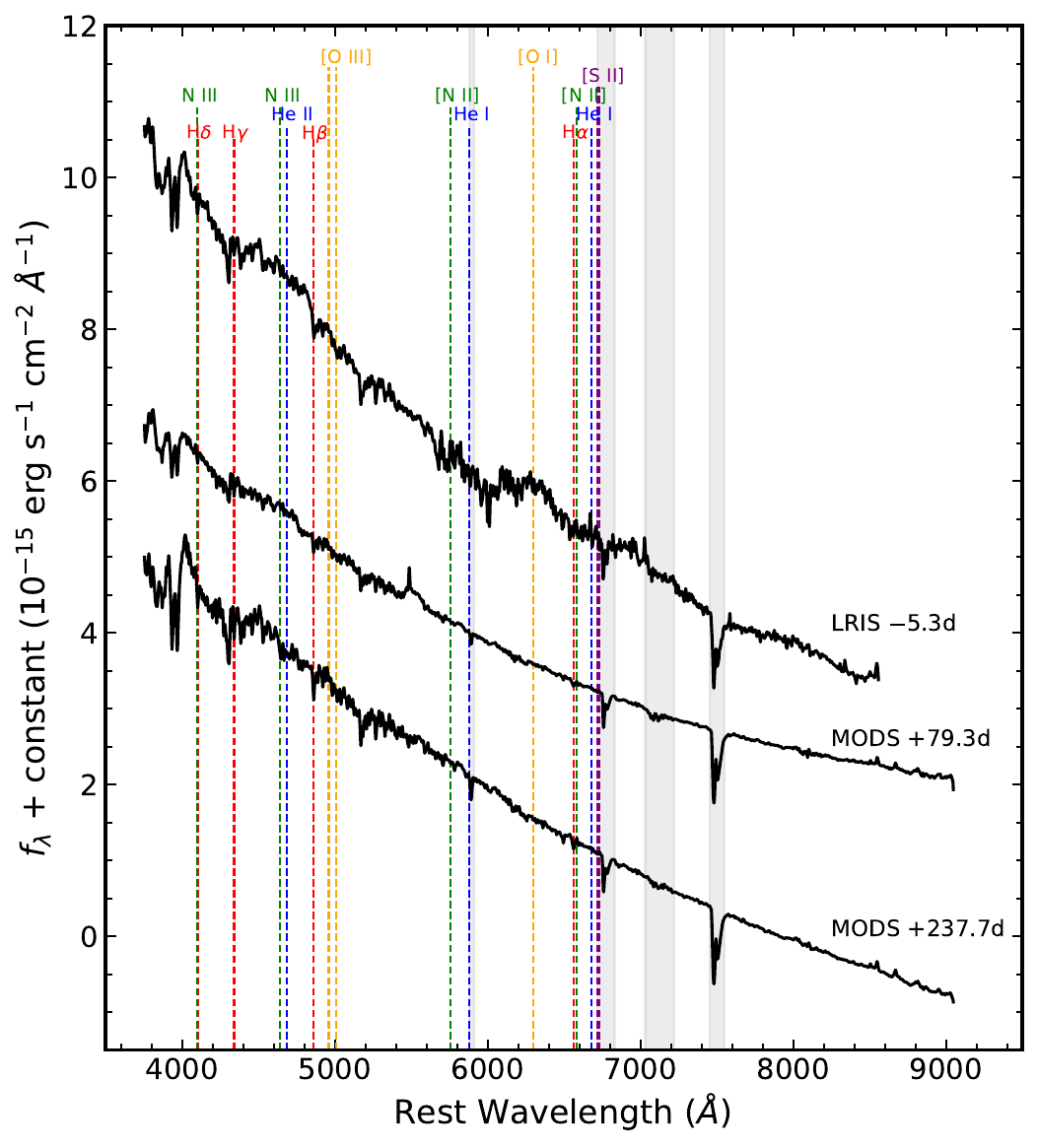}\hfill
 \caption{High S/N host-subtracted spectra from LRIS and MODS at three phases throughout the evolution of ASASSN-20hx. The features near 6300 \AA\ and 5500 \AA\ in the LRIS and MODS spectra respectively are reduction artifacts that were accentuated by the blackbody scaling and host subtraction procedures. The vertical gray bands mark atmospheric telluric features. The vertical lines mark spectral features common in TDEs and AGNs, with hydrogen lines in red, helium lines in blue, nitrogen lines in green, sulfur lines in purple, and oxygen lines in orange.}
 \label{fig:high_snr}
\end{figure*}

Figure \ref{fig:high_snr} shows three high S/N host-subtracted spectra from various phases in the evolution of ASASSN-20hx. None of the expected emission lines are visible in these deep spectra. The broad feature at $\sim 6300$ \AA\ in the LRIS spectrum is not present in a SPRAT spectrum taken the same day, suggesting it is a reduction artifact caused by known issues with the red channel of LRIS at the time the spectrum was obtained. This artifact was accentuated by the blackbody scaling and host subtraction procedures. In none of these spectra do we see evidence for lines consistent with a TDE, AGN, or SN origin.

\begin{figure}
\centering
 \includegraphics[width=0.48\textwidth]{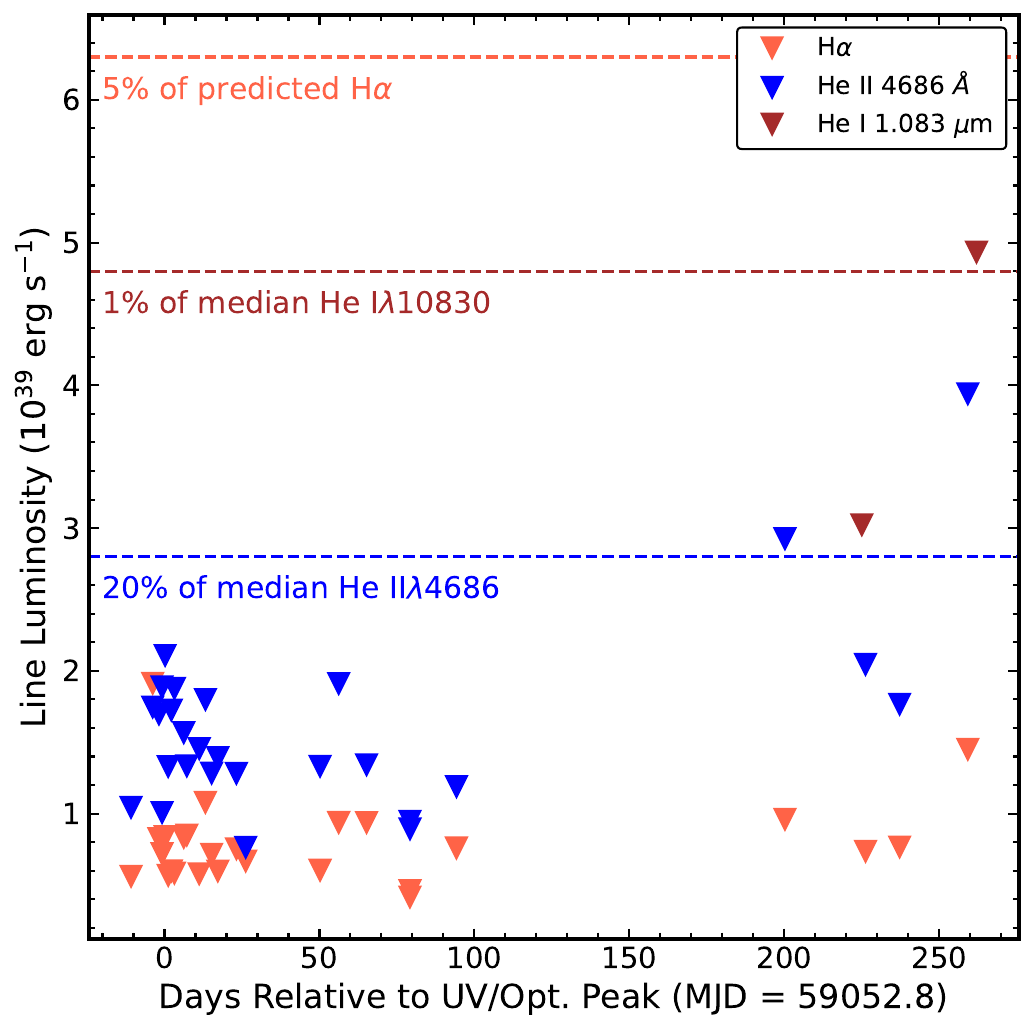}\hfill
 \caption{5$\sigma$ limits on the H$\alpha$, \ion{He}{2}$ \lambda4686$, and \ion{He}{1} $\lambda10830$ emission from ASASSN-20hx. The dashed red line indicates 5\% of the expected H$\alpha$ emission for an AGN with the X-ray luminosity of ASASSN-20hx using the correlation of \citet{shi10}. The dashed blue line is 20\% of the median \ion{He}{2} $\lambda4686$ luminosity from the sample of \citet{bentz10}. The dashed brown line is 1\% of the median \ion{He}{1} $\lambda10830$ luminosity from the sample of \citet{landt08}. In the case of H$\alpha$ and \ion{He}{1} $\lambda10830$, the limits are at the level of a few percent or lower and at roughly the ten percent level for \ion{He}{2} $\lambda4686$.}
 \label{fig:line_lims}
\end{figure}

We computed limits on the existence of H$\alpha$, \ion{He}{2} $\lambda4686$, and \ion{He}{1} $\lambda10830$ from our optical and NIR spectra to quantify the weakness of any line features. To do this, we followed the procedure of \citet{leonard01} and \citet{tucker20}, assuming a AGN-like line width of 2000 km s$^{-1}$ and obtained 5$\sigma$ flux limits as
\begin{equation}
F(5\sigma) = 5C_{\lambda}\Delta I \sqrt{W_{line}\Delta X}
\end{equation}
where $C_{\lambda}$ is the continuum flux at wavelength $\lambda$, $\Delta I$ is the RMS scatter around a normalised continuum, $W_{line}$ is the width of the line profile, and $\Delta X$ is the pixel scale of the spectrum. The results are shown in Figure \ref{fig:line_lims}. The majority of the limits for both H$\alpha$ and \ion{He}{2} $\lambda4686$ are below $\sim 2 \times 10^{39}$ erg s$^{-1}$, one to two orders of magnitude less than the line emission typically seen in AGNs or TDEs \citep[e.g.,][]{bentz10, gezari12b, holoien19a, neustadt20}. Additionally, the limits on H$\alpha$ lie at a few percent of the H$\alpha$ emission predicted from the correlation between X-ray luminosity and H$\alpha$ luminosity in AGNs from \citet{shi10}. The limits on H$\alpha$ luminosity are also much less than TDEs with H emission \citep[e.g.][]{holoien14b, holoien19b, hinkle21a}. We also find that the NIR \ion{He}{1} $\lambda10830$ emission is less than a percent of what is seen in known broad-line AGN. While the comparison samples used here are heterogeneous, they all illustrate that the limits on the line emission from ASASSN-20hx are very strong, with none of these lines approaching typical levels. The few epochs with larger limits on line luminosity have noticeable noise in their spectra or occur at crossover regions in the spectrograph on which they were taken.

\begin{figure*}
\centering
 \includegraphics[width=0.95\textwidth]{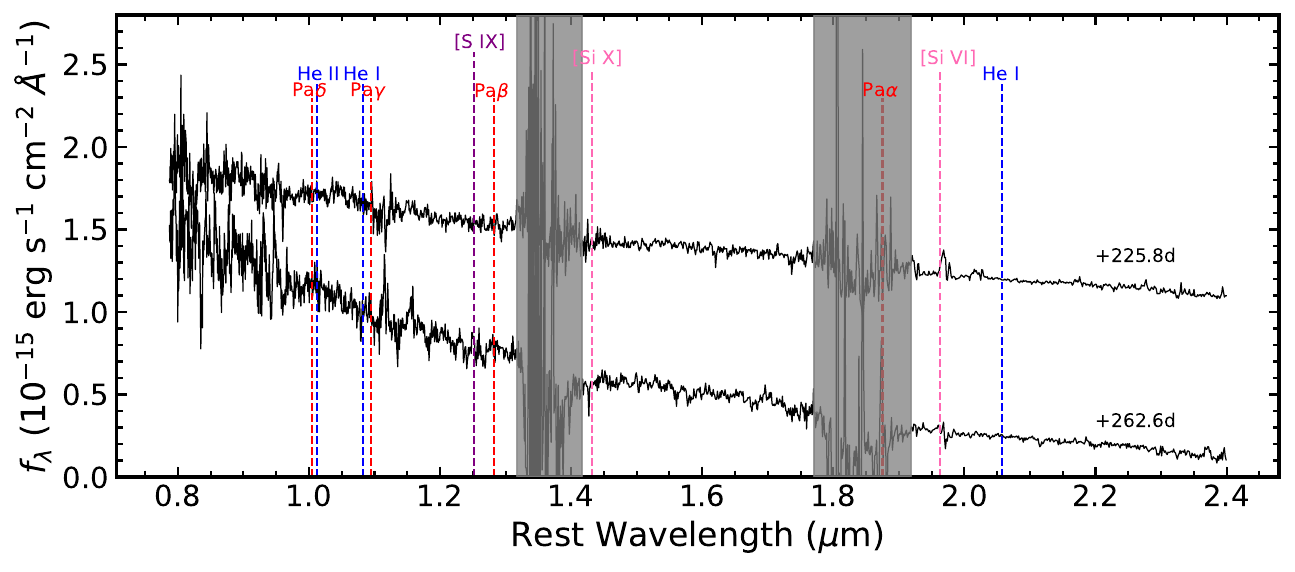}
 \caption{NIR spectra of ASASSN-20hx obtained with IRTF 225.2 days after peak. The vertical gray bands mark strong atmospheric telluric features. The vertical lines mark spectral features common in AGN, with hydrogen lines in red and helium lines in blue. Coronal silicon and sulfur lines are shown in pink and purple respectively. The spectra have been binned to $\sim 5$ \AA \ bins.}
 \label{fig:nir_spec}
\end{figure*}

Given the lack of features in the optical spectra, we obtained a near infrared spectrum, shown in Figure \ref{fig:nir_spec}, to search for lines. Many of the H and He lines are more isolated in the NIR than in the optical. As with the optical spectra there is no clear evidence for any H or He emission that might suggest an AGN origin for this transient. We also search for high ionization coronal lines \citep[e.g.,][]{lamperti17}, but found no evidence for any of the lines commonly seen in AGN.

\subsection{Spectral Energy Distribution} \label{sed}

\begin{figure}
\centering
 \includegraphics[width=0.48\textwidth]{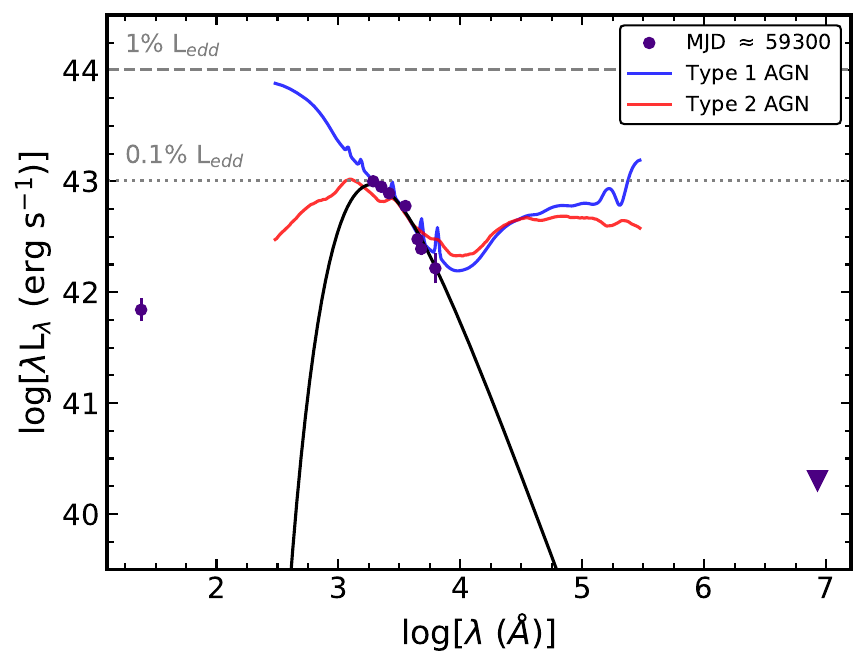}\hfill
 \caption{Spectral energy distribution of ASASSN-20hx at MJD $\sim$ 53900, at roughly 250 days after peak. The black line is the corresponding best-fitting blackbody model at this phase. The downward facing triangle shows the 3$\sigma$ upper-limit on the sub-mm emission at the measured frequency. The horizontal gray dashed and dotted lines show 1\% and 0.1\% of the Eddington luminosity for a SMBH of $10^{7.9}$ \msun respectively. The blue and red lines show typical SEDs of Type 1 and Type 2 AGNs respectively from \citet{assef10} normalized to the ASASSN-20hx blackbody fit at the $U$-band.}
 \label{fig:SED}
\end{figure}

The spectral energy distribution of ASASSN-20hx is shown in Figure \ref{fig:SED}. These data are taken at a phase of roughly 250 days after peak, where we have X-ray, UV/optical, and sub-mm constraints on the transient emission. The SED shows that the UV/optical emission from ASASSN-20hx is dominant over the X-ray emission and that the total emission is far below the Eddington luminosity for the expected black hole mass of the host galaxy. While the dominance of the UV/optical emission and the good fit of a blackbody SED is similar to known TDEs \citep[e.g.,][]{holoien14b, holoien16a, vanvelzen21}, it is also qualitatively similar to unobscured AGNs over this wavelength range \citep[e.g.,][]{assef10}.

\begin{figure}
\centering
 \includegraphics[width=0.48\textwidth]{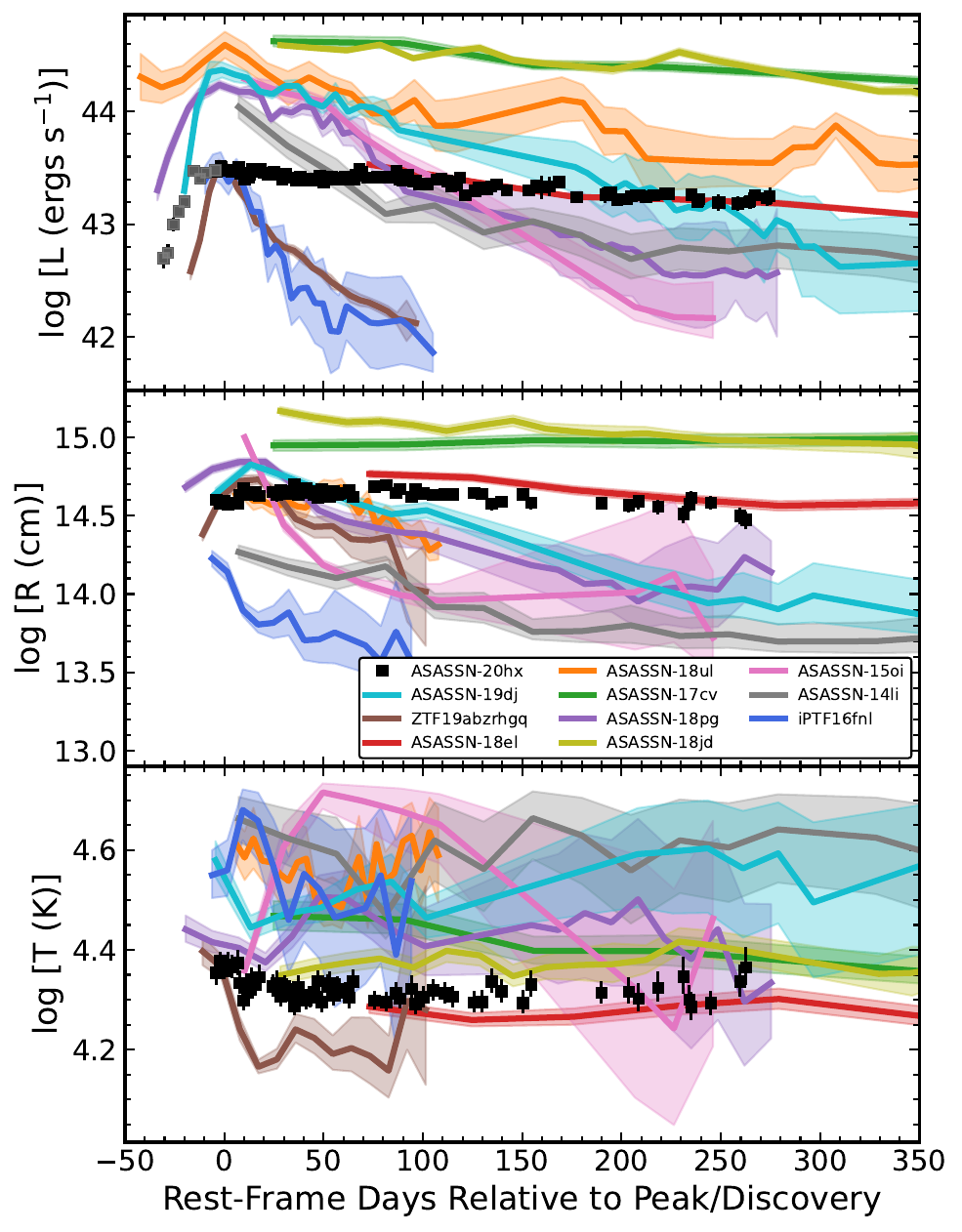}\hfill
 \caption{Evolution of the UV/optical blackbody luminosity (top panel), radius (middle panel), and temperature (bottom panel) for ASASSN-20hx (black squares) and a comparison sample of TDEs and ANTs. The lines are smoothed over the individual epochs by linearly interpolating to a time-series with the same length as the original coverage, but with half the number of points. Time is in rest-frame days relative to the peak luminosity for the objects discovered prior to peak (ASASSN-19dj, ASASSN-18ul, ZTF19abzrhgq, ASASSN-18pg, and iPTF16fnl), and relative to discovery for those which were not (ASASSN-18el, ASASSN-17cv, ASASSN-18jd, ASASSN-15oi, and ASASSN-14li). The gray squares for ASASSN-20hx indicate where data has been bolometrically corrected using the ASAS-SN $g$-band light curve assuming the temperature from the first \swift epoch was constant.}
 \label{fig:BB_fit}
\end{figure}

Given the dominance of the UV/optical emission from ASASSN-20hx, we fit it as a blackbody, similar to many previous TDEs and nuclear flares \citep{hinkle21b}. Figure \ref{fig:BB_fit} shows the blackbody model fits in terms of luminosity, radius, and temperature for ASASSN-20hx compared to the well-studied TDEs ASASSN-19dj \citep[light blue line;][]{hinkle21a}, ZTF19abzrhgq \citep[AT2019qiz; brown line;][]{nicholl20}, ASASSN-18ul \citep[AT2018fyk, orange line;][Payne et al., in preparation]{wevers19}, ASASSN-18pg \citep[purple line;][]{holoien20}, ASASSN-15oi \citep[pink line;][]{holoien18a}, ASASSN-14li \citep[gray line;][]{brown17a}, and iPTF16fnl \citep[dark blue line;][]{blagorodnova17, brown18} and to the nuclear transients ASASSN-18el \citep[red line;][]{trakhtenbrot19b}, ASASSN-17cv \citep[AT2017bgt, green line;][]{trakhtenbrot19a}, and ASASSN-18jd \citep[gold line;][]{neustadt20}. ASASSN-20hx is among the least luminous of the well-studied nuclear transients, with a peak luminosity of $(3.15 \pm 0.04) \times 10^{43} \text{ erg s}^{-1}$. Using the SMBH mass computed from scaling relations, the peak luminosity corresponds an Eddington ratio of $\sim 3 \times 10^{-3}$, roughly 100 times less than typical TDEs \citep{wevers17} and consistent with the lower end of the distribution of optically selected AGNs \citep{kauffmann09}. This peak luminosity is comparable to the TDEs ZTF19abzrhgq \citep[AT2019qiz; ][]{nicholl20} and iPTF16fnl \citep[][]{blagorodnova17, brown18}. Unlike these faint TDEs, the decline in bolometric luminosity of ASASSN-20hx is quite slow. The decline slope of the bolometric luminosity of ASASSN-20hx appears similar to the other ANTs in our sample, which are all flatter than the TDE comparison sample.

The blackbody radius of ASASSN-20hx is large compared to the well-studied TDEs, but it is similar to the ANTs in the comparison sample. Unlike many TDEs with a well-sampled late-time evolution \citep[e.g.,][]{vanvelzen21, hinkle20a, hinkle21a} the blackbody radius of ASASSN-20hx does not decrease significantly at late times. This relatively flat late-time radius evolution is similar to other AGN-related nuclear flares \citep[e.g.,][]{hinkle21b}.

The effective temperature of ASASSN-20hx is moderately cool compared to other TDEs and ANTs, on the order of $\sim 21,000$ K. Nonetheless, the temperature of ASASSN-20hx is similar to some nuclear flares like ASASSN-18el. Interestingly, the blackbody temperature of ASASSN-20hx is most similar to TDEs exhibiting strong H emission \citep{hung20, holoien19a, holoien19b} although ASASSN-20hx does not display H emission. Like the TDEs and nuclear flares in our comparison sample, ASASSN-20hx displays a flat temperature throughout its evolution, with only a small decrease soon after peak.

\begin{figure}
\centering
 \includegraphics[width=0.48\textwidth]{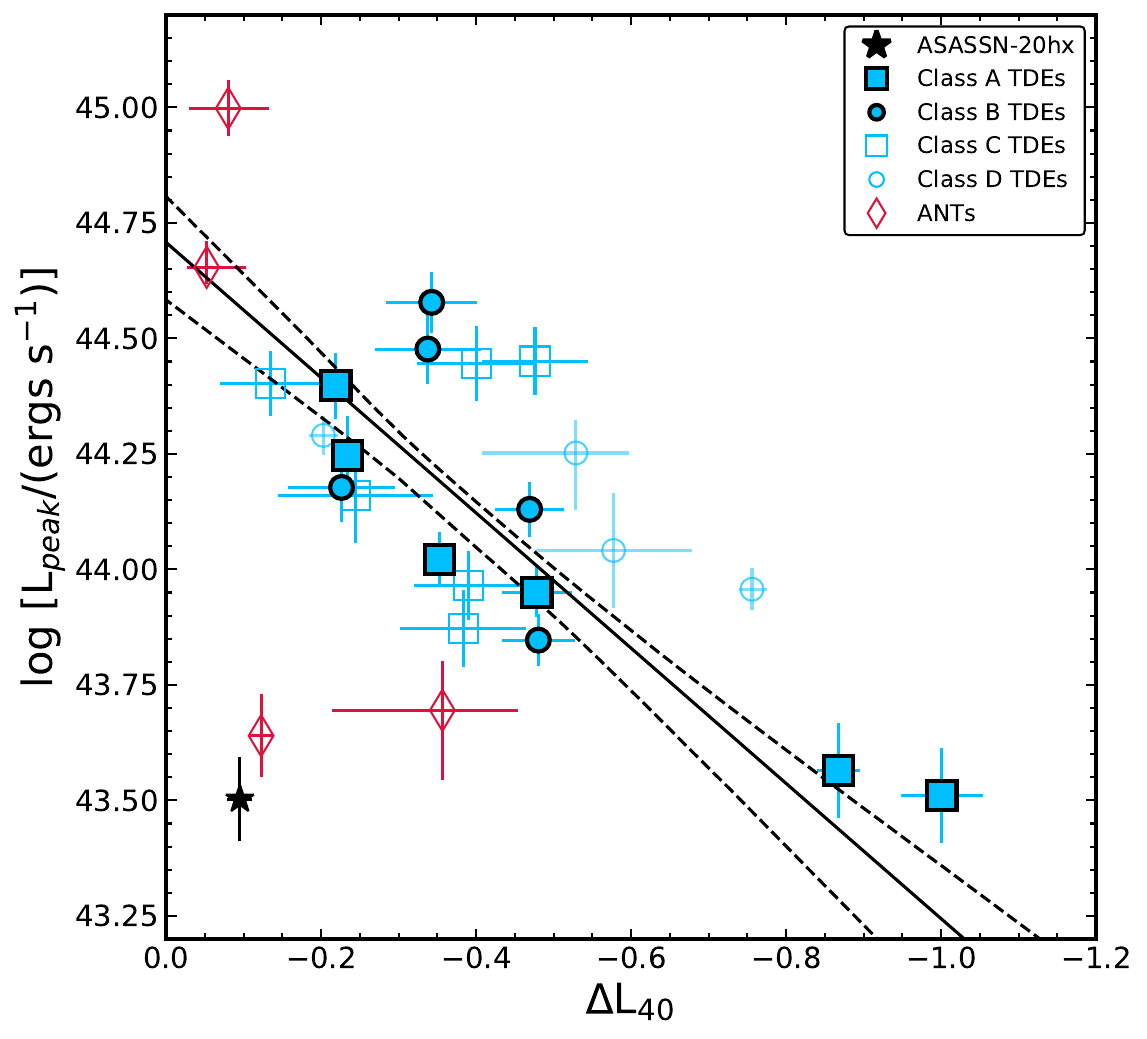}
 \caption{The peak luminosity versus the decline rate of various TDEs and other nuclear outbursts. The decline rate $\Delta$L$_{40}$ is defined as the difference between the log of the peak luminosity and the log of the luminosity at 40 days after peak. The squares and circles correspond to the various classes of objects in \citet{hinkle20a, hinkle21b} and the diamonds are events interpreted as AGN flares. The black solid line is the best-fit line for the TDEs and the dashed black lines are the allowed range of uncertainty from the best-fit line. ASASSN-20hx lies well below the TDE relationship and is more consistent with the decline rates of the non-TDE nuclear transients.}
 \label{fig:deltaL40}
\end{figure}

Finally, with bolometric light curves, we can compare the light curve evolution of ASASSN-20hx to other well-studied TDEs and nuclear flares using the method of \citet{hinkle20a}. By measuring the peak UV/optical luminosity and comparing it to the decline rate, we can see where ASASSN-20hx lies relative to TDEs of similar luminosities as shown in Figure \ref{fig:deltaL40}. The first striking point is that ASASSN-20hx is less luminous than any of the TDEs in this sample. For the fundamental plane distance of 55.6 Mpc \citep{saulder06}, this luminosity difference would be even stronger. Unlike the two faintest TDEs to date, ASASSN-20hx does not fade quickly, declining by only $\sim 0.1$ dex in luminosity over the first forty days after peak. While much less luminous, this decline rate is similar to the other ANTs. The fact that ASASSN-20hx lies far from the known TDE relationship and occupies a region of decline rate space similar to AGN-related transients strongly suggests that it is not a normal TDE. As suggested by its similarities to ASASSN-18el in Figure \ref{fig:BB_fit}, ASASSN-20hx lies close to this ANT in Figure \ref{fig:deltaL40}.

\subsection{X-rays} \label{sec:xray}

\begin{deluxetable}{ccccc}
\tablewidth{240pt}
\tabletypesize{\footnotesize}
\tablecaption{X-ray Photon Indices}
\tablehead{
\colhead{MJD} &
\colhead{MJD Range} &
\colhead{$\Gamma$} & 
\colhead{$\Gamma$ Error} &
\colhead{Satellite} }
\startdata
59051.38 & 2.23 & 2.29 & 0.17 & \swift \\
59055.14 & --- & 2.51 & 0.13 & NICER \\
59057.08 &  --- & 2.69 & 0.12 & NICER \\
\ldots & \ldots & \ldots & \ldots & \ldots \\
59177.90 & 12.11 & 2.28 & 0.23 & \swift \\
59230.25 & 35.17 & 2.18 & 0.30 & \swift \\
59303.98 & 28.92 & 2.05 & 0.30 & \swift \\
\enddata 
\tablecomments{X-ray photon indices from the fits to the individual NICER spectra or the stacked \swift spectra. The MJD range indicates the length in time over which the \swift spectra were stacked.} 
\label{tab:xray_spec} 
\end{deluxetable}

\begin{figure*}
\centering
 \includegraphics[width=1.0\textwidth]{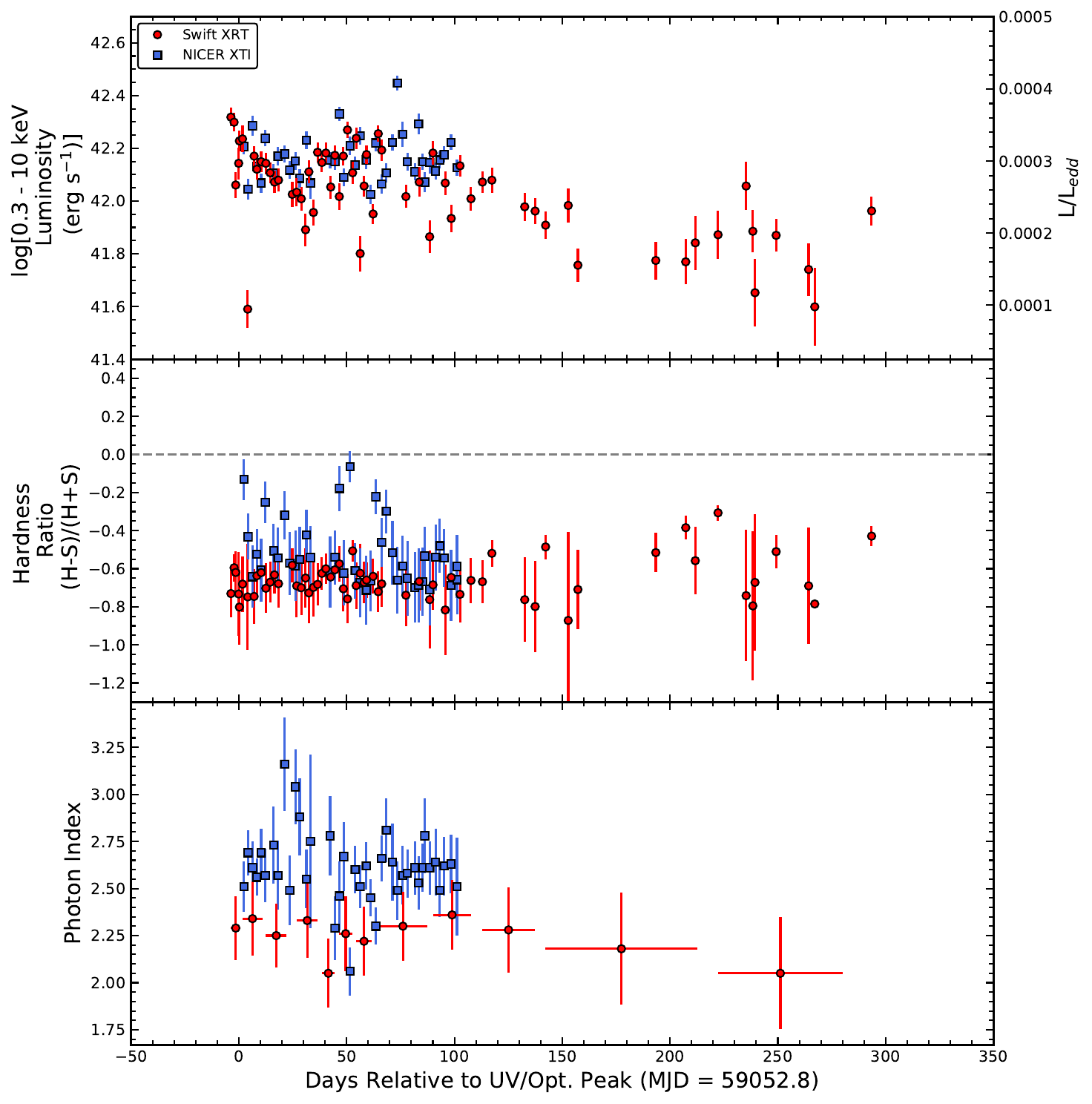}
 \caption{X-ray luminosity (top panel), hardness ratio (middle panel), and photon index (bottom panel) of ASASSN-20hx measured with \textit{Swift} (red circles) and \textit{NICER} (blue squares). We define hard counts H as the number of counts in the 2-10 keV range and soft counts S are the number of counts in the 0.3-2 keV range, with a gray dashed line marking zero. The hardness ratio is defined as (H$-$S)/(H+S).}
 \label{fig:all_xray}
\end{figure*}

While the host galaxy of ASASSN-20hx showed weak X-ray emission prior to the current outburst, the X-ray emission associated with the transient is roughly an order of magnitude higher. In Figure \ref{fig:all_xray} (top panel), we show the X-ray light curve as derived from the \textit{Swift} and \textit{NICER} observations. To estimate the X-ray luminosity, we converted the extracted count rate into flux using WebPIMMS\footnote{\url{https://heasarc.gsfc.nasa.gov/cgi-bin/Tools/w3pimms/w3pimms.pl}} and assumed an absorbed power-law model with the average photon indices derived from our fits. 

We analyzed the \textit{Swift} and \textit{NICER} spectra using the X-ray spectral fitting program \textsc{XSPEC} version 12.10.1f \citep{arnaud96}, and chi-squared statistics. While we fit the \textit{NICER} spectra individually, it was necessary to stack the \swift observations to get adequate S/N for spectral fits. Each of the spectra was well-fit by an absorbed power law. The fits did not require additional absorption beyond the Galactic contribution, so all of the column densities were frozen at the Galactic value of $2.0 \times 10^{20}$ cm$^{-2}$. We show the resulting photon indices from these spectral fits in the bottom panel of Figure \ref{fig:all_xray}. 

Throughout the evolution of ASASSN-20hx, the merged \swift spectra and individual NICER spectra are best fit by an absorbed power-law model. In a few \textit{NICER} spectra, there was marginal evidence for a blackbody component with kT $\sim 0.035$ keV and a radius of $\sim 5 \times 10^{11}$ cm. However, as this additional component is only needed for a small number of spectra and at moderate significance, we do not explore it further. In Table \ref{tab:xray_spec} we give the best-fit photon indices from our spectral fits.

For \textit{Swift} the average photon index was $\Gamma = 2.24$ and for \textit{NICER} it was $\Gamma = 2.61$, although they are largely consistent within the uncertainties for the overlapping data. The presence of power-law X-ray emission is unlike many X-ray luminous TDEs \citep[e.g., ASASSN-14li, ASASSN-15oi, ASASSN-19dj;][]{brown17a, holoien18a, kara18, hinkle21a}, and is more consistent with an AGN scenario. The average photon index derived for ASASSN-20hx is typical of unobscured AGNs \citep{ricci17}. 

The first X-ray observation of ASASSN-20hx was taken using the \textit{Swift} XRT roughly 3 days before the peak UV/optical emission (MJD = 59052.4). During this observation, ASASSN-20hx showed strong X-ray emission at a luminosity of $\sim2\times10^{42}$ erg s$^{-1}$. This is higher than many TDEs in the early phases of their evolution \citep{auchettl17}, but is similar to or less than the X-ray luminosities of other nuclear flares that have been suggested to be associated with AGN activity, such as ASASSN-18jd \citep{neustadt20}, ASASSN-17cv \citep[AT2017bgt; ][]{trakhtenbrot19a}, and several transients in Narrow-Line Seyfert 1 galaxies \citep{frederick20}.

Unlike many other nuclear transients, such as ASASSN-18jd \citep{neustadt20}, ASASSN-18ul \citep[AT2018fyk; ][]{wevers19}, and ASASSN-15oi \citep{gezari17}, the X-ray emission of ASASSN-20hx showed little variability over the first $\sim$100 days after peak, varying between $\sim10^{42.0} - 10^{42.2}$ erg s$^{-1}$. Unlike typical X-ray luminous TDEs like ASASSN-14li \citep{holoien16a} and ZTF19aapreis \citep[AT2019dsg;][]{vanvelzen21}, the X-ray luminosity of ASASSN-20hx did not decrease significantly over the first $\sim$200 days after peak. Using only the X-ray emission of ASASSN-20hx, we find an Eddington ratio of $\sim 3 \times 10^{-4}$, roughly 5 orders of magnitude higher than the Eddington ratio of Sgr A* computed from X-ray emission during flaring activity \citep{zhang17}. 

The second panel of Figure \ref{fig:all_xray} shows the evolution of the X-ray hardness ratio HR = (H$-$S)/(H+S), defining the hard (H) and soft (S) bands as 2-10 keV and 0.3-2 keV respectively. ASASSN-20hx shows little variability in its hardness, with a HR range of roughly $-0.8$ to $-0.3$, excluding some individual \textit{NICER} epochs with hardness ratios around zero. As the X-ray emission begins to fade towards $\sim 200$ days after peak, the hardness ratios increase slightly, but still remain below zero. The hardness ratios of ASASSN-20hx are generally consistent with AGNs \citep{auchettl18} and are slightly harder than most TDEs \citep{auchettl17, auchettl18, hinkle21a}.

\begin{figure}
\centering
 \includegraphics[width=0.48\textwidth]{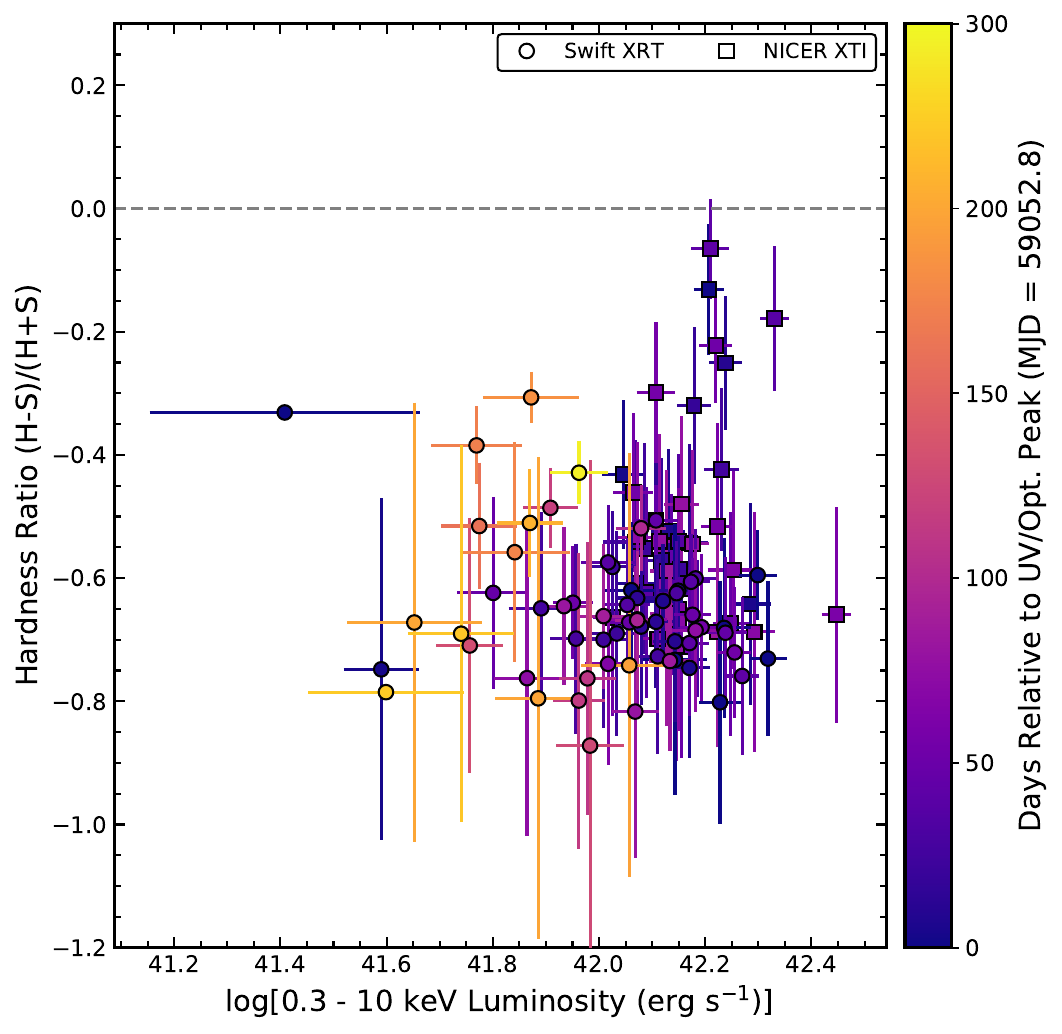}
 \caption{Hardness ratio as a function of X-ray luminosity as measured by \textit{Swift} (circles) and \textit{NICER} (squares) for the detections only. While the overall trend is small, as ASASSN-20hx becomes brighter, the X-ray emission becomes softer. This behavior is similar to that seen in X-ray bright AGNs \citep{auchettl18}. The color bar on the right indicates the phase relative to the UV/optical peak, with darker colors indicating earlier times.}
 \label{fig:hr_lum}
\end{figure}

In the X-ray evolution of ASASSN-20hx, the HR and the X-ray luminosity evolve slowly with time. They seem to follow a weak inverse relationship, where the X-ray emission becomes harder as the luminosity of the source fades. This relationship is shown in Figure \ref{fig:hr_lum}. Albeit much weaker for ASASSN-20hx, this evolution is consistent with what is seen in highly variable, X-ray bright AGNs \citep[][]{auchettl18} and some TDEs like ASASSN-19dj \citep{hinkle21a}.

\begin{figure*}
\centering
 \includegraphics[width=0.95\textwidth]{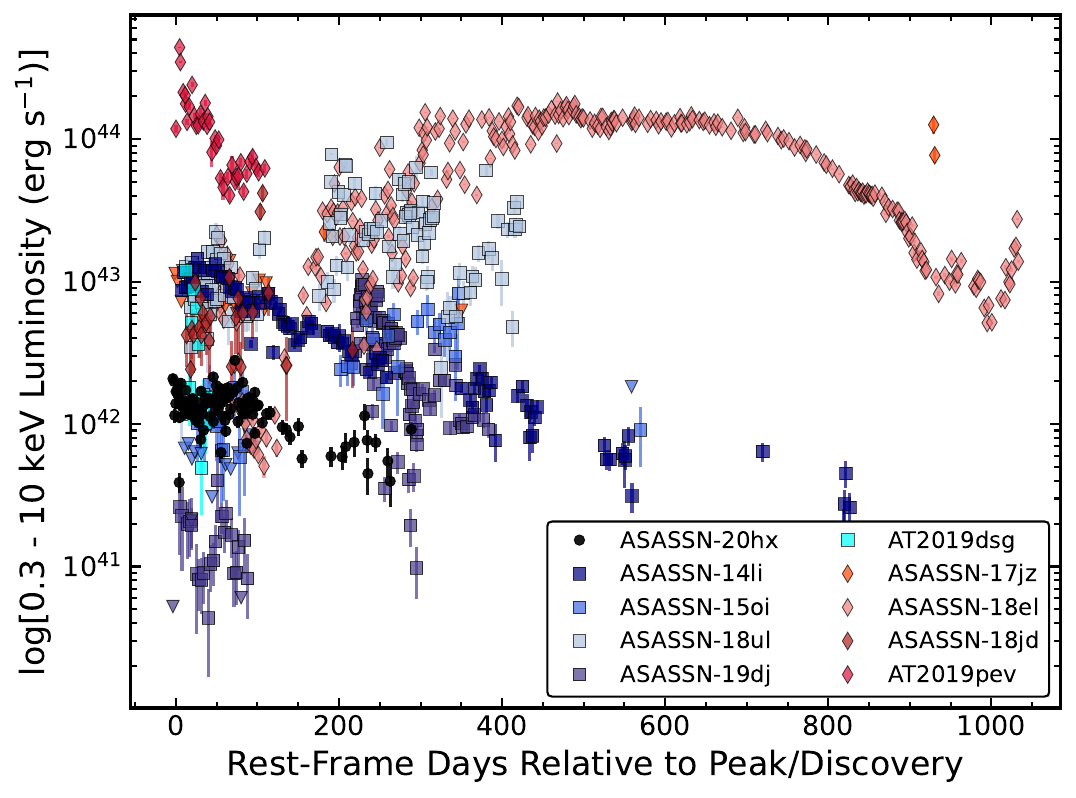}
 \caption{Evolution of the 0.3 - 10 keV X-ray luminosities for ASASSN-20hx (black squares) and a comparison sample of TDEs (blue squares) and ANTs (red diamonds). Time is in rest-frame days relative to the peak UV/optical luminosity for the objects discovered prior to peak (ASASSN-18ul, ASASSN-19dj, AT2019dsg, ASASSN-17jz, and AT2019pev) and relative to discovery for those which were not (ASASSN-18el, ASASSN-18jd, ASASSN-15oi, and ASASSN-14li).}
 \label{fig:xray_comparison}
\end{figure*}

Figure \ref{fig:xray_comparison} shows the X-ray luminosity of ASASSN-20hx compared to the well-studied TDEs ASASSN-14li \citep{brown17a}, ASASSN-15oi \citep{holoien18a}, ASASSN-18ul \citep[AT2018fyk; ][Payne et al., in preparation]{wevers19}, ASASSN-19dj \citep{hinkle21a}, and ZTF19aapreis \citep[AT2019dsg; ][Auchettl et al., in preparation]{vanvelzen21} and to the nuclear transients ASASSN-17jz \citep[][]{holoien21}, ASASSN-18el \citep[][Hinkle et al., in preparation]{trakhtenbrot19b}, ASASSN-18jd \citep[][]{neustadt20}, and ZTF19abvgxrq \citep[AT2019pev; ][Auchettl et al., in preparation]{frederick20}. The X-ray luminosity of ASASSN-20hx near UV/optical peak is below the median luminosity seen for X-ray bright TDEs and ANTs, but still higher than some like ASASSN-19dj. Many TDEs and ANTs show a late-time re-brightening period in their X-ray light curves, particularly notable for ASASSN-15oi, ASASSN-18el, and ASASSN-19dj. Over the roughly 275 days studied here for ASASSN-20hx, we see no evidence for such a feature. Indeed, the X-ray evolution is largely dissimilar to all other objects in our comparison sample. ASASSN-14li and ZTF19aapreis/AT2019dsg both show monotonically declining X-ray light curves somewhat like ASASSN-20hx, but at much higher peak luminosities and different decline rates. Much like the other multi-wavelength properties of ASASSN-20hx, the X-rays do not neatly agree with observations of known TDEs or AGN flares.

\section{Discussion}\label{disc}

ASASSN-20hx is a unique transient and perhaps the most clearly ambiguous of the growing class of ANTs. Much like other objects in this class, such as ASASSN-18jd \citep{neustadt20}, ASASSN-18el \citep{trakhtenbrot19b}, ASASSN-17cv \citep[AT2017bgt; ][]{trakhtenbrot19a}, and ZTF19aaiqmgl \citep[AT2019avd;][]{frederick20, malyali21}, ASASSN-20hx shares some characteristics with both known TDEs and AGN. In Figures \ref{fig:BB_fit} and \ref{fig:deltaL40} we have compared the blackbody evolution and the decline rate to various well-studied TDEs and ANTs. Throughout the rest of this section, we will examine the properties of ASASSN-20hx as compared to other types of nuclear transients.

\subsection{TDE-like features of ASASSN-20hx}

One of the key features of ASASSN-20hx that made it a strong candidate for a TDE based on its early evolution \citep{hinkle20_atel} was the strong UV emission and blue colors. While less luminous than most TDEs, the UV emission from ASASSN-20hx was still substantial, with $M_{UV} \sim -18.6$ mag near peak. The SED of ASASSN-20hx is well-fit by a blackbody, similar to many TDEs \citep[e.g.,][]{hinkle20a, vanvelzen21}. The blackbody temperature of ASASSN-20hx is roughly 21,000 K throughout its evolution, which is cooler than many TDEs, but still consistent with the larger population. Like most TDEs, the temperature of ASASSN-20hx stays relatively constant over time.

The peak luminosity of ASASSN-20hx of $(3.15 \pm 0.04) \times 10^{43}$ erg s$^{-1}$ is very low for a TDE. Only iPTF16fnl \citep{blagorodnova17} and ZTF19abzrhgq \citep[AT2019qiz; ][]{nicholl20} have similar luminosities, but both of these TDEs fade quickly after their peaks (see Fig. \ref{fig:deltaL40}). While at a similar peak luminosity, ASASSN-20hx fades very slowly after peak. In contrast, the decline rate of ASASSN-20hx is slower than any TDE included in the sample of \citet{hinkle20a}.

The UV/optical light curve evolution of ASASSN-20hx is similar to some TDEs in that the overall shape is quite smooth, although there are hints of weak short-term variability in the light curves. Additionally it is worth noting that some TDEs, like PS18kh \citep{holoien19b, vanvelzen19}, do show variability and moderate rebrightening episodes. The long-lived, flat UV/optical light curve without a significant initial decay is similar to the predictions for TDEs on more massive SMBHs \citep{mummery21a}.

ASASSN-20hx is one of the few nuclear transients where the slope of the initial flux increase can be measured. All of the TDEs for which such a fit has been possible (ASASSN-19bt, ASASSN-19dj, and ZTF19abzrhgq) have a flux $\propto t^2$ initial rise \citep{holoien19c, hinkle21a, nicholl20}, while the flux of ASASSN-20hx rises linearly with time. However, the transient ASASSN-14ko, which is presently intepreted as a repeating partial TDE \citep{payne21, payne21b, tucker21} also shows a linear rise in its TESS light curve. The overall time for ASASSN-20hx to reach its peak UV/optical emission is similar to typical TDEs.

Next, we can compare the host galaxy of ASASSN-20hx to other TDE hosts. As shown in Figure \ref{fig:ew_bpt}, the host galaxy has less evidence of recently formed stars than many TDE hosts. As with most TDE hosts, the archival optical emission is non-variable and shows no evidence of previous outbursts (see Figs. \ref{fig:long_lc} and \ref{fig:TESS_lc}). The SMBH mass derived from scaling relations for the host galaxy is $10^{7.9}$ \msun, which is higher than most TDEs and very close to the point where the observable TDE rate begins to decrease rapidly \citep{kochanek16b, vanvelzen18, wevers19a}.

The lack of broad spectral lines is odd for the TDE scenario. While various lines are seen at differing strengths in TDEs \citep[e.g.,][]{leloudas19, vanvelzen21}, H or He lines are almost always seen. In only one case, PS1-11af \citep{chornock14}, has a TDE not exhibited significant line emission. However, spectra were only obtained for PS1-11af near peak emission so we do not know if lines appeared later. It is possible to delay the production of strong lines \citep{roth18}, which has been observed for the TDE ASASSN-19dj \citep[e.g.,][]{hinkle21a}. This could explain the lack of emission lines, but lines typically appear relatively quickly after peak ($\sim 30$ days), whereas ASASSN-20hx is still devoid of emission lines after a year. 

The X-ray emission of ASASSN-20hx is also unlike any other TDE, starting with the fact that it is best fit by a power law rather than a blackbody. This is uncommon for TDE X-ray spectra \citep{auchettl17}. However, \citet{mummery21b} suggest that TDEs on more massive SMBHs should show non-thermal X-ray spectra and evolve more slowly, much like the observed behavior of ASASSN-20hx. While the hardness ratios of ASASSN-20hx are somewhat soft, they are still harder than many other X-ray luminous TDEs like ASASSN-14li \citep{holoien16a} and ASASSN-19dj \citep{hinkle21a}. 

\subsection{AGN-like features of ASASSN-20hx}

The first characteristic of ASASSN-20hx that suggests a potential AGN origin is the archival \swift XRT X-ray detection. At a luminosity of $(2.2 \pm 1.4)  \times 10^{41} \text{ erg } \text{ s}^{-1}$, with little star formation in the host, the X-rays are likely to originate from an LLAGN \citep[][]{tozzi06, marchesi16, liu17, ricci17}. Given the host stellar mass, this X-ray luminosity is roughly two orders of magnitude higher than expected from low-mass X-ray binaries \citep{gilfanov04}.

Indeed, the X-ray properties of ASASSN-20hx are generally consistent with an AGN. The power-law spectrum with a photon index of $\Gamma \sim 2.3$ is typical of an unobscured AGN. The hardness ratios are soft, but within the range seen for other AGNs \citep{auchettl18}. Additionally, the Eddington ratios measured from the X-ray emission and the peak UV/optical luminosity are much more consistent with low-luminosity AGNs \citep[e.g.,][]{ricci17} than a typical TDE \citep[e.g.,][]{mockler19}.

While the SED of ASASSN-20hx is TDE-like in terms of its blackbody nature and persistent hot temperature, the temporal evolution of ASASSN-20hx in terms of blackbody properties is similar to other ANTs that have been argued to be of AGN origin in the literature.  In Figure \ref{fig:BB_fit}, the luminosity decline of ASASSN-20hx is similar to ASASSN-18el \citep{trakhtenbrot19b} and ASASSN-17cv \citep[AT2017bgt; ][]{trakhtenbrot19a}. This is further supported by the narrow range in the $\Delta L_{40}$ decline rates seen in the ANTs in Figure \ref{fig:deltaL40}. Similar to the other AGN-related transients, ASASSN-20hx has a relatively large effective blackbody radius. In fact, whereas most TDEs have shrinking blackbody radii over time \citep{hinkle20a, vanvelzen21}, ASASSN-20hx and the other ANTs have roughly constant effective radii for over a year. All of the nuclear transients shown in Figure \ref{fig:BB_fit} are hot, but ASASSN-20hx and the other non-TDE transients have among the coolest blackbody temperatures. This is perhaps due to the more massive black holes where these transients tend to occur \citep[e.g.,][]{shakura73}.

Save for the archival X-ray detection, the host galaxy of ASASSN-20hx does not have many clear signs of AGN activity. Figure \ref{fig:wise} shows the WISE color-color selection of \citet{assef13} along with well-studied ANTs and TDEs. Most ANTs and TDEs lie far from the shaded gray region that indicates AGN selection but the host galaxy of ASASSN-20hx has the bluest $W1 - W2$ color. The host galaxy of ASASSN-20hx is classified as an AGN in the [\ion{N}{2}]/H$\alpha$ diagram, its classification is ambiguous in the [\ion{S}{2}]/H$\alpha$ diagram and it would not be classified as an AGN in the WHAN diagram. This lack of consistent classification makes it unlikely that the host galaxy of ASASSN-20hx was a strong AGN.

The lack of H, He, N, O, and S optical and NIR emission lines is also difficult to reconcile with an AGN origin. In the spectra of ASASSN-20hx, these lines are all very weak or absent entirely. Normal AGN all have emission lines \citep[e.g.,][]{ho08} which typically increase in strength during a flare \citep[e.g,][]{trakhtenbrot19a, frederick19, frederick20}. Even if we were to imagine a scenario where low-ionization lines were not produced during the ASASSN-20hx flare, we might expect the high level of UV and X-ray photons to produce the coronal lines often seen in AGNs \citep{lamperti17}. However, as shown in Figure \ref{fig:nir_spec}, none of these commonly seen lines are present in the NIR spectra of ASASSN-20hx, further challenging a typical AGN picture.

One class of objects with AGN-like X-ray emission without corresponding optical AGN lines are the XBONGs \citep[X-ray Bright Optically Normal Galaxies; ][]{moran02, smith14}, although these typically do not lack emission lines entirely. If the lack of emission lines seen for ASASSN-20hx is simply the result of missing gas in the nucleus, continued follow-up over several years may reveal narrow lines at later times. Additionally, blazars show featureless continua with no emission lines, but here we would likely expect radio emission if the jet is not significantly self-absorbed.

\subsection{Comparing TDE and AGN Scenarios}

\begin{deluxetable}{ccc}
\tablewidth{240pt}
\tabletypesize{\footnotesize}
\tablecaption{AGN vs. TDE Properties of ASASSN-20hx}
\tablehead{
\colhead{Property} &
\colhead{ASASSN-20hx obs.} &
\colhead{AGN/TDE like} }
\startdata
log(M$_{BH}$)$ < 8$ \msun & Yes & TDE \\
$\sigma_{H\beta} < 2000$ km s$^{-1}$ & N/A & N/A \\
\ion{Fe}{2} Emission & No & TDE \\
\ion{O}{3} $/ H\beta <$ 3 & Yes & AGN \\
$\Delta (g - r) \sim 0$ & Yes & TDE \\
UV-bright & Yes & TDE \\
X-ray Power Law & Yes & AGN \\
$W1 - W2 > 0.7$ mag & No & TDE \\
Rebrightening & No & TDE \\
\enddata 
\tablecomments{Properties to differentiate between TDEs and AGN flares, inspired by the classification scheme of \citet{frederick20}. Based on whether or not ASASSN-20hx shows a given property, the final column indicates if this behavior is more AGN-like or TDE-like.} 
\label{tab:agn_tde_comp} 
\end{deluxetable}

Using the framework of \citet{frederick20}, we can compare the various characteristics of ASASSN-20hx and assign them to a more AGN-like or more TDE-like classification. This classification method was developed for transients occurring in Narrow-line Seyfert 1 galaxies, but the basic principles can be applied more broadly. First, the SMBH mass for ASASSN-20hx is below $10^8$ \msun, and thus is consistent with a TDE. As there is no measured hydrogen emission during outburst, we cannot compute a line width to discriminate between typical TDE and AGN line widths. We do not observe \ion{Fe}{2} emission from ASASSN-20hx, consistent with a TDE as opposed to an AGN. The UV/optical colors are roughly constant over time, which is consistent with a TDE, as is the UV luminosity of the event. The existence of a power-law-like X-ray emission at a relatively hard photon index is more consistent with an AGN than a TDE. As previously mentioned, the WISE colors of the host are inconsistent with a strong AGN, although these data are of the quiescent host emission. Finally, while the slow decrease in the luminosity of ASASSN-20hx over time with no significant rebrightenings is consistent with a TDE, the very long overall timescale for fading is more consistent with known transients in AGNs rather than typical TDEs. 

In addition to the events studied in \citet[][]{frederick20}, there have been several claims in the literature of TDEs occurring in galaxies hosting an AGN. These include PS16dtm \citep{blanchard17} and ASASSN-18el \citep{ricci20}. In these cases, the host galaxies were known AGNs and the transient emission and evolution were argued to be consistent with a TDE superimposed on top of the existing AGN variability. Such a scenario is possible for ASASSN-20hx and could help explain the existence of TDE-like UV/optical emission with AGN-like X-ray emission. Nonetheless, such a scenario is speculative and difficult to prove definitively. In addition, the optical and NIR spectra are still inconsistent with this picture, so such an explanation does not allow us to neatly explain the full range of observed properties for ASASSN-20hx.

\section{Summary}\label{summary}

We have presented the discovery of the ANT ASASSN-20hx and detailed the multi-wavelength photometric and spectroscopic follow-up data we have obtained on this event. The key properties of ASASSN-20hx are as follows:

\begin{itemize}
    \item The host galaxy of ASASSN-20hx did not host a strong AGN prior to the transient.
    \item The rise in flux of ASASSN-20hx is well-constrained by TESS photometry with a power-law slope of $\alpha = 1.05 \pm 0.06$, which is shallower than the $\alpha \sim 2$ previously seen in TDEs.
    \item The UV/optical SED of ASASSN-20hx is well-fit by a blackbody with a roughly constant temperature of $\sim 21,000$ K and a peak luminosity of $3.2 \times 10^{43}$ erg s$^{-1}$. The evolution of the blackbody properties is qualitatively consistent with previous TDEs and ANTs, but not typical AGNs.
    \item The optical and NIR spectra of ASASSN-20hx are devoid of emission lines, which is problematic for either an AGN or TDE explanation for this transient. 
    \item The host galaxy of ASASSN-20hx was detected as an X-ray source prior to the ASASSN-20hx flare. The X-ray emission associated with ASASSN-20hx is roughly an order of magnitude higher and well-fit by a power law with $\Gamma \sim 2.3-2.6$, consistent with AGN-like emission.
\end{itemize}

Many of the observed properties of ASASSN-20hx can be reconciled through the TDE or AGN explanation, or in some cases both. However, the lack of emission lines proves difficult to explain in either case. Nonetheless as ASASSN-20hx continues to evolve and we continue to add more ANTs to our growing sample, we will be able to better understand the full range of behaviors occurring in galactic nuclei.

\section*{Acknowledgements}
We thank the referee for helpful comments and suggestions that have improved the quality of this manuscript. We thank Connor Auge, Richard Mushotzky, and Andrew Mummery for helpful discussions. We thank J.D. Armstrong for help scheduling LCOGT follow-up photometry. We thank the \swift PI, the Observation Duty Scientists, and the science planners for promptly approving and executing our \swift observations. Additionally we thank the \textit{NICER} PI and the science planners for promptly approving and executing our \textit{NICER} observations.

We thank the Las Cumbres Observatory and its staff for its continuing support of the ASAS-SN project. ASAS-SN is supported by the Gordon and Betty Moore Foundation through grant GBMF5490 to the Ohio State University, and NSF grants AST-1515927 and AST-1908570. Development of ASAS-SN has been supported by NSF grant AST-0908816, the Mt. Cuba Astronomical Foundation, the Center for Cosmology  and AstroParticle Physics at the Ohio State University, the Chinese Academy of Sciences South America Center for Astronomy (CAS- SACA), the Villum Foundation, and George Skestos. 

JTH was supported by NASA grant 80NSSC21K0136 throughout this work. BJS, CSK, and KZS are supported by NSF grant AST-1907570/AST-1908952. BJS is also supported by NSF grants AST-1920392 and AST-1911074. CSK and KZS are supported by NSF grant AST-181440. M.A.T acknowledges support from the DOE CSGF through grant DE-SC0019323. Support for JLP is provided in part by FONDECYT through the grant 1191038 and by the Ministry of Economy, Development, and Tourism's Millennium Science Initiative through grant IC120009, awarded to The Millennium Institute of Astrophysics, MAS. TAT is supported in part by NASA grant 80NSSC20K0531. PJV is supported by the National Science Foundation Graduate Research Fellowship Program Under Grant No. DGE-1343012

Parts of this research were supported by the Australian Research Council Centre of Excellence for All Sky Astrophysics in 3 Dimensions (ASTRO 3D), through project number CE170100013.

This publication makes use of data products from the Wide-field Infrared Survey Explorer, which is a joint project of the University of California, Los Angeles, and the Jet Propulsion Laboratory/California Institute of Technology, funded by the National Aeronautics and Space Administration.

This paper includes data collected by the TESS mission. Funding for the TESS mission is provided by the NASA's Science Mission Directorate.

This paper contains data obtained at the Wendelstein Observatory of the
Ludwig-Maximilians University Munich.

The Liverpool Telescope is operated on the island of La Palma by Liverpool John Moores University in the Spanish Observatorio del Roque de los Muchachos of the Instituto de Astrofisica de Canarias with financial support from the UK Science and Technology Facilities Council.

The LBT is an international collaboration among institutions in the United States, Italy and Germany. LBT Corporation partners are as follows: The University of Arizona on behalf of the Arizona Board of Regents; Istituto Nazionale di Astrofisica, Italy; LBT Beteiligungsgesellschaft, Germany, representing the Max-Planck Society, The Leibniz Institute for Astrophysics Potsdam, and Heidelberg University; The Ohio State University, representing OSU, University of Notre Dame, University of Minnesota, and University of Virginia.

Some of the data presented herein were obtained at the W. M. Keck Observatory, which is operated as a scientific partnership among the California Institute of Technology, the University of California and the National Aeronautics and Space Administration. The Observatory was made possible by the generous financial support of the W. M. Keck Foundation.

Visiting Astronomer at the Infrared Telescope Facility, which is operated by the University of Hawaii under contract 80HQTR19D0030 with the National Aeronautics and Space Administration.

This work is based on observations made by ASAS-SN, UH88, JCMT, and Keck. We wish to extend our special thanks to those of Hawaiian ancestry on whose sacred mountains of Maunakea  and Haleakal\=a, we are privileged to be guests. Without their generous hospitality, the observations presented herein would not have been possible.

\begin{deluxetable*}{cccccc}
\tablewidth{240pt}
\tabletypesize{\footnotesize}
\tablecaption{Spectroscopic Observations of ASASSN-20hx}
\tablehead{
\colhead{MJD} &
\colhead{UTC Date} &
\colhead{Telescope} &
\colhead{Instrument} & 
\colhead{Rest Wavelength Range (\AA)} &
\colhead{Exposure Time (s)} }
\startdata
59043.0 & 2020 July 13.0 & Liverpool 2-m Telescope & SPRAT & $3954-7863$ & 1$\times$900 \\
59047.5 & 2020 July 17.5 & Keck I 10-m Telescope & LRIS & $3024-8557$ & 1$\times$600 + 1$\times$1200 \\
59052.0 & 2020 July 22 & Liverpool 2-m Telescope & SPRAT & $3981-7863$ & 1$\times$800 \\
59052.5 & 2020 July 22.5 & Keck I 10-m Telescope & LRIS & $3006-8557$ & 2$\times$950 \\
59053.0 & 2020 July 23.0 & Liverpool 2-m Telescope & SPRAT & $3954-7863$ & 1$\times$800 \\
59054.0 & 2020 July 24.0 & Liverpool 2-m Telescope & SPRAT & $3954-7863$ & 1$\times$800 \\
59055.0 & 2020 July 25.0 & Liverpool 2-m Telescope & SPRAT & $3954-7863$ & 1$\times$800 \\
59056.0 & 2020 July 26.0 & Liverpool 2-m Telescope & SPRAT & $3954-7863$ & 1$\times$800 \\
59057.0 & 2020 July 27.0 & Liverpool 2-m Telescope & SPRAT & $3954-7863$ & 1$\times$800 \\
59060.0 & 2020 July 30.0 & Liverpool 2-m Telescope & SPRAT & $3954-7863$ & 1$\times$800 \\
59060.9 & 2020 July 30.9 & Liverpool 2-m Telescope & SPRAT & $3954-7863$ & 1$\times$800 \\
59065.0 & 2020 August 4.0 & Liverpool 2-m Telescope & SPRAT & $3954-7863$ & 1$\times$800 \\
59066.9 & 2020 August 5.9 & Liverpool 2-m Telescope & SPRAT & $3954-7863$ & 1$\times$1200 \\
59069.0 & 2020 August 8.0 & Liverpool 2-m Telescope & SPRAT & $3954-7863$ & 1$\times$1200 \\
59071.0 & 2020 August 10.0 & Liverpool 2-m Telescope & SPRAT & $3954-7863$ & 1$\times$1200 \\
59076.9 & 2020 August 15.9 & Liverpool 2-m Telescope & SPRAT & $3954-7863$ & 1$\times$1200 \\
59079.3 & 2020 August 18.3 & Keck I 10-m Telescope & LRIS & $3340-9901$ & $3\times1050$ (blue), $3\times970$ (red) \\
59103.2 & 2020 September 11 & Large Binocular Telescope 8.4-m & MODS & $3443-9048$ & 3$\times$1200 \\
59109.8 & 2020 September 17.8 & Liverpool 2-m Telescope & SPRAT & $3954-7863$ & 2$\times$1200 \\
59118.8 & 2020 September 26.8 & Liverpool 2-m Telescope & SPRAT & $3954-7863$ & 2$\times$1200 \\
59132.1 & 2020 October 10 & Large Binocular Telescope 8.4-m & MODS & $3443-9048$ & 3$\times$1200 \\
59132.8 & 2020 October 10.8 & Liverpool 2-m Telescope & SPRAT & $3954-7863$ & 2$\times$1200 \\
59147.8 & 2020 October 25.8 & Liverpool 2-m Telescope & SPRAT & $3954-7863$ & 2$\times$1200 \\
59253.6 & 2021 February 8.6 & University of Hawaii 88-in Telescope & SNIFS & $3347-8950$ & 1$\times$2700 \\
59278.6 & 2021 March 5.6 & NASA Infrared Telescope Facility & SpeX & $7871-23993$ & 24$\times$50 \\
59279.6 & 2021 March 6.6  & University of Hawaii 88-in Telescope & SNIFS & $3347-8950$ & 1$\times$2700 \\
59290.5 & 2021 March 17 & Large Binocular Telescope 8.4-m & MODS & $3443-9048$ & 4$\times$900s \\
59312.6 & 2021 April 8.6 & University of Hawaii 88-in Telescope & SNIFS & $4040-8950$ & 1$\times$2700 \\
59315.4 & 2021 April 11.4 & NASA Infrared Telescope Facility & SpeX & $7871-23993$ & 8$\times$200 \\
\enddata 
\tablecomments{Modified Julian Day, calendar date, telescope, instrument, wavelength range, and exposure time for each of the spectroscopic observations obtained of ASASSN-20hx for the initial classification and during our follow-up campaign. For some of the Keck LRIS spectra the red and blue sides have slightly different exposure times due to the very different readout times of the two chips. The two LT/SPRAT spectra taken on the same day as the Keck/LRIS and LBT/MODS spectra are not shown in Fig. \ref{fig:opt_spec}.} 
\label{tab:spectra_log} 
\end{deluxetable*}

\bibliography{bibliography}
\bibliographystyle{aasjournal}


\label{lastpage}
\end{document}